\documentclass{elsarticle}

\usepackage[percent]{overpic}

\usepackage{rotating}
\usepackage{mathrsfs}
\usepackage{lineno,hyperref}
\usepackage{tikz}
\usepackage[export]{adjustbox}
\usepackage{amsmath}
\usepackage{mathtools}
\usepackage{geometry,tikz}
\usetikzlibrary{arrows,matrix,positioning,fit}
\usepackage{amssymb}
\usepackage{physics}
\usepackage{float}
\usepackage{subcaption}
\usepackage{array}
\usepackage{multirow}
\usepackage{makecell}
\usepackage{relsize}
\usepackage{ulem}
\usepackage{enumitem}
\usepackage[super]{nth}
\usepackage{relsize}
\modulolinenumbers[5]

\usepackage[ruled,vlined,linesnumbered]{algorithm2e}
\SetArgSty{textnormal} 

\journal{XXX}

\newcommand{\fig}[1]{Fig.~\ref{#1}}
\newcommand{\eq}[1]{Eq.~(\ref{#1})}

\newcommand{\tab}[1]{Tab.~\ref{#1}}

\newcommand{\sgn}{\text{sgn}}
\DeclareCaptionFormat{cont}{#1 #2#3\par}

\newcommand{\derisk}{DeRisk}
\newcommand{\un}{u_n} 
 
\newcommand{\azLag}{\dot{w}}
\newcommand{\sx}{s} 
\newcommand{\de}{\dot{\eta}} 
\newcommand{\sun}{\overline{\sigma}_{\un}}
\newcommand{\ssx}{\overline{\sigma}_{\sx}}
\newcommand{\hs}{H_{\rm s}}
\newcommand{\tz}{T_{\rm z}}
\newcommand{\depth}{h}
\newcommand{\hsnom}{{H_{\rm s}}^{\rm (nom)}}
\newcommand{\Fx}{K_x}
\newcommand{\Fy}{K_y}
\newcommand{\Ix}{I_x}
\newcommand{\Iy}{I_y}
\newcommand{\Iun}{I_{\un}}
\newcommand{\Fxmax}{{\Fx}^{\rm max}}
\newcommand{\Fymax}{{\Fy}^{\rm max}}
\newcommand{\twe}{t_{\rm we}}
\newcommand{\tadim}{\tilde{t}}
\newcommand{\Xv}{\boldsymbol{X}}
\newcommand{\Uv}{\boldsymbol{U}}
\newcommand{\Yv}{\boldsymbol{Y}}
\newcommand{\Yi}{Y_i}
\newcommand{\Yri}{\boldsymbol{Y}_{\backslash i}}
\newcommand{\Yrk}{\boldsymbol{Y}_{\backslash k}}
\newcommand{\Zi}{\boldsymbol{Z}_{|i}}
\newcommand{\Zk}{\boldsymbol{Z}_{|k}}
\newcommand{\ai}{\boldsymbol{a}_{|i}}
\newcommand{\bi}{\boldsymbol{b}_{|i}}
\newcommand{\aai}{\boldsymbol{\alpha}_{|i}}
\newcommand{\bbi}{\boldsymbol{\beta}_{|i}}
\newcommand{\mi}{\boldsymbol{\mu}_{|i}}
\newcommand{\si}{\boldsymbol{\sigma}_{|i}}

\newcommand{\yth}{y_{\rm th}}
\newcommand{\eHT}{\varepsilon_{\rm HT}}
\newcommand{\nHT}{N_{\rm HT}}
\newcommand{\nMT}{N_{\rm MT}}
\newcommand{\nT}{N_{\rm tail}}
\newcommand{\mTail}{M_{\rm tail}}

\newcommand{\pHT}{P_{\rm HT}}
\newcommand{\seqHT}{\tau_{\rm HT}}
\newcommand{\rBulk}{\mathcal{R}_{\rm bulk}}
\newcommand{\mBulk}{\mathcal{M}_{\rm bulk}}
\newcommand{\rHTall}{\mathcal{R}_{\rm HT}}
\newcommand{\npca}{N_{\rm pc}}
\newcommand{\nfeat}{d}

\newcommand{\mHT}{\mathcal{M}_i}
\newcommand{\mHTk}{\mathcal{M}_k}
\newcommand{\rHT}{\mathcal{R}_i}

\newcommand{\nobs}{n}
\newcommand{\nobsHT}{{n_{\rm HT}}}

\newcommand{\FLap}{F_{\rm L}}
\newcommand{\SGPD}{S_{\rm G}}
\newcommand{\CDFXi}{F_{i}}
\newcommand{\dquant}{\Delta Q _{\Fxmax}^{\rm (nb-wb)}}
\newcommand{\qnnb}{Q_{\Fxmax}^{\rm (nb)}}
\newcommand{\qnwb}{Q_{\Fxmax}^{\rm (wb)}}
\makeatletter
\def\ps@pprintTitle{%
  \let\@oddhead\@empty
  \let\@evenhead\@empty
  \let\@oddfoot\@empty
  \let\@evenfoot\@empty}
\makeatother

\begin{document}


\begin{frontmatter}

\title{
Data-driven modeling of multivariate stochastic trajectories -- Application to water waves
}


\author[enstaAddress]{Romain Hasco\"{e}t\corref{cor1}}
\cortext[cor1]{Corresponding author}
\ead{romain.hascoet@ensta.fr}

\address[enstaAddress]{IRDL, UMR CNRS 6027, ENSTA, Institut Polytechnique de Paris, \\ Bretagne INP, Univ. Brest, Univ. Bretagne Sud, Brest, France}

\begin{abstract}

A data-driven methodology is proposed to model the distribution of multivariate stochastic trajectories from an observed sample.
As a first step, each trajectory in the sample is reduced to a vector of features by means of Functional Principal Component Analysis. 
Next, the joint distribution of features is modeled using (i) a non-parametric vine copula approach for the bulk of the distribution, and (ii) the conditional modeling framework of Heffernan \& Tawn (2004) for the multivariate tail.
The method is applied to the modeling of water waves.
The dataset used is the \derisk{} database, which consists of numerical simulations of water waves. The analysis is restricted to the portion of the wave period between the free-surface zero-upcrossing and the wave crest.
The kinematic variables considered are the free-surface slope, the normal component of the fluid velocity at the free surface, and the vertical Lagrangian acceleration of the fluid at the free surface. The stochastic trajectories of these three variables are modeled jointly.
The vertical Lagrangian acceleration of the fluid is employed to enforce a wave-breaking filter in the stochastic model.
The number of hyperparameters in the stochastic framework is reduced to three, and a stepwise calibration strategy is proposed for their adjustment.
The capabilities of the model are illustrated by predicting the distributions of selected response variables and by generating synthetic trajectories.

\end{abstract}

\begin{keyword}
water wave \sep
sea state \sep
Functional PCA  \sep
extreme value theory \sep 
stochastic trajectory  \sep
vine copula
\end{keyword}

\end{frontmatter}

\section{Introduction}

In ocean and coastal environments, marine structures are subjected to loads induced by water waves -- see, e.g., \cite{naess_1992, faltinsen_1993, faltinsen_1995, bachynski_2014, renaud_2025}.
These loads can induce seakeeping motions and cause structural damage, either from a single extreme event exceeding the ultimate strength limit or from fatigue due to repeated loading -- see e.g. \cite{jha_2000, li_2013, storhaug_2014, hascoet_2025}.
Accurate evaluation of these loads is therefore crucial when designing marine structures.
Ocean surface waves — and thus the loads they generate — are inherently random.
To represent this randomness, the free-surface elevation and the associated fluid kinematics are commonly modeled as a multivariate random field depending on spatial and temporal coordinates.
To make the problem tractable, it is typically analyzed at two distinct scales -- see, e.g., \cite{kinsman_1984, tucker_2001, ochi_2005, holthuijsen_2007}. 
At short timescales (from a few tens of minutes to a few hours) and over short spatial scales (from a few hundred meters to a few tens of kilometers), the random field representing water waves is assumed to be stationary and homogeneous.
These locally stationary conditions define a \textit{sea state}.
Then, to account for the stochastic evolution of water waves over larger timescales and spatial scales, the wave field is modeled as a succession of stationary sea states adjoining in space and time.

Under the simplest modeling assumptions, the wave field is described by the linear wave model, which allows the associated random field to be treated as Gaussian and often renders the problem analytically tractable (see, e.g., \cite{longuet_1953, lindgren_1982, azais_2005, hascoet_2021}).
Other model-driven approaches incorporate second-order nonlinear effects, enabling the reproduction of certain non-Gaussian features observed in the statistical properties of water waves (see, e.g., \cite{longuet_1964, naess_1985, langley_1987, hascoet_2022}).
Beyond second order, incorporating nonlinear effects into the stochastic modeling of water waves becomes highly challenging, if not impractical, with a purely model-driven approach.
Then, one must rely on experiments -- whether laboratory wave tank tests, in situ measurements at sea, or numerical simulations -- to build a database on which a data-driven model can be formulated and calibrated. 
To the best of our knowledge, the use of data-driven approaches to model individual water waves has received limited attention in the literature.

In this study, we propose a data-driven approach to model the joint dynamics of random processes characterizing water waves.
The database employed is the \derisk{} database \cite{pierella_2021}, which is populated with numerical simulations performed using a potential flow model.
The kinematic variables considered to illustrate the approach -- namely the normal component of the fluid velocity at the free surface, the free surface slope, and the vertical Lagrangian acceleration of the fluid at the free surface -- were selected for their direct relevance to the design of marine structures.
To address the proposed problem, we first employ Functional Principal Component Analysis to reduce the stochastic trajectories to low-dimensional feature vectors.
Next, the joint distribution of the features is modeled, addressing both the bulk (i.e., the central part) and the tail of the distribution.
For the bulk of the distribution, a non-parametric vine-copula approach is employed, while for the multivariate tail, the semi-parametric method of Heffernan \& Tawn \cite{heffernan_2004} is implemented.
The approach is assessed by generating synthetic water-wave trajectories from the fitted model, computing the distributions of selected response variables, and comparing them with the corresponding empirical distributions from the \derisk{} database. 

The paper is structured as follows. 
Section \ref{sec_framework} establishes the study framework, providing a brief description of the \derisk{} database and introducing the wave-related processes to be considered.
Section \ref{sec_stocha_trajs_reduction} describes the strategy employed to reduce the trajectories (i.e., realizations) of the random processes to low-dimensional feature vectors.
Section \ref{sec_joint_distribution} details how the distribution of the feature vector is modeled: the bulk of the distribution is captured using a vine copula approach (Section \ref{subsubsec_bulk_model}), while the multivariate tail is modeled using the conditional approach of Heffernan \& Tawn (2004) \cite{heffernan_2004} (Section \ref{subsec_HT_model}).
Section \ref{seq_strategy_hyperparams} proposes a stepwise strategy for selecting the three main hyperparameters of the stochastic model.
Section \ref{sec_restults} presents a selection of results obtained with the proposed approach. 
The adequacy of the model is assessed by comparing simulated and true trajectories, as well as by comparing the empirical and predicted distributions of selected response variables.
The paper concludes with a summary and discussion in Section \ref{sec_discussion}.

\section{The dataset and considered variables}
\label{sec_framework}

\subsection{The \derisk{} database: selected case studies}
\label{subsec_select_case_studies}

The database used in the present study is the \derisk{} database \cite{pierella_2021}, which is populated with simulations of fully nonlinear irregular waves based on the potential flow code OceanWave3D \cite{engsigkarup_2009}.
Different sea states are simulated for various combinations of significant wave height, $\hs$, mean wave period, $\tz$, and water depth, $\depth$. All sea states in the \derisk{} database are unidirectional.
For each sea state, multiple simulation runs were performed (typically $8$ runs). To ensure that each run represents a different realization of the same sea state, a distinct set of random phases was assigned to the components of the wave spectrum specified at the entry zone of the simulation domain.
To prevent instabilities caused by near-breaking waves, OceanWave3D implements a breaking filter that limits the downward Lagrangian acceleration of the fluid to a fraction $\beta$ of the gravitational acceleration, $g$:
\begin{equation}
\label{eq_breaking_filter}
- \azLag < \beta g \, , 
\end{equation}
where $\azLag$ is the vertical Lagrangian acceleration, defined as positive when pointing upward.
By comparing the extreme wave elevations and wave kinematics predicted by the numerical model with those measured in wave-tank experiments (see \cite{pierella_2021}), the value $\beta = 0.5$ was selected to generate the \derisk{} database.

\def\scaleF{0.5}
\def\fI{./s}

\begin{figure}[t!]
\begin{center}
\begin{tabular}{c} 
	\includegraphics[width=\scaleF\textwidth]{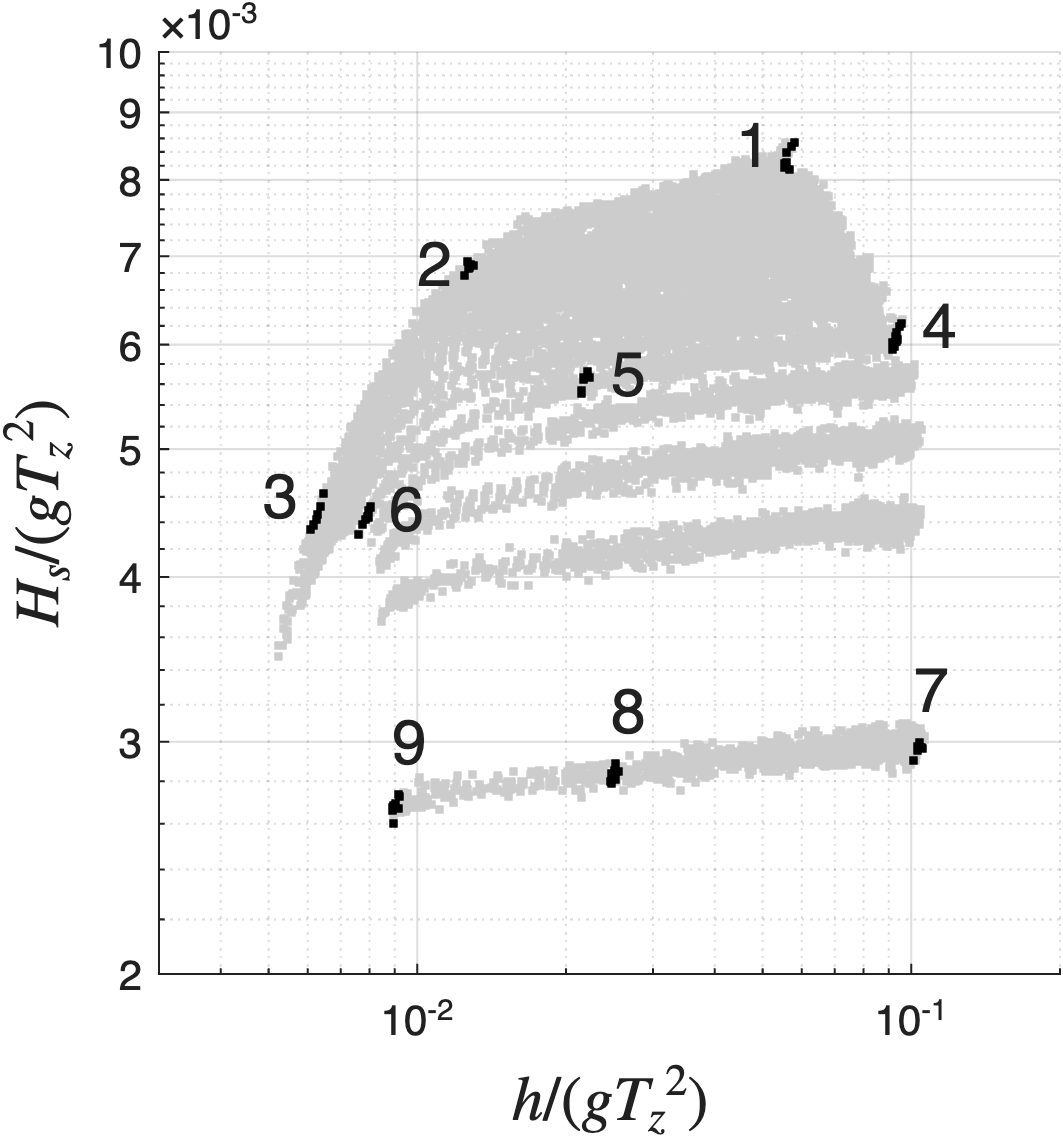}
\end{tabular}        
\end{center}
\caption{
Parameter space of sea states covered by the \derisk{} simulations. Gray dots indicate individual simulation runs, while clusters of black dots, labeled with numbers, denote the sea states analyzed in this study (see \tab{tab_Derisk_sims}).
}
\label{fig_Derisk_sims}
\end{figure}

\begin{table}[h!]
\begin{center}
    \begin{tabular}{| c | c | c | c | c | c |}
    \hline
    $\#$ sea state & $\hsnom \ [{\rm m}]$ & $\depth \ [{\rm m}]$ & $\hs\ [{\rm m}]$ & $\tz\ [{\rm s}]$ & $T_{\rm sim} \ [{\rm s}]$ \\ 
    \hline
    $1$ & $22.52$ & $125.0$ & $18.4$ & $15.0$ & $1.16 \ 10^5$ \\ 
    \hline
    $2$ & $22.52$ & $30.55$ & $16.5$ & $15.6$ & $1.16 \ 10^5$ \\ 
    \hline
    $3$ & $22.52$ & $15.17$ & $10.8$ & $15.7$ & $1.16 \ 10^5$ \\ 
    \hline    
    $4$ & $10.14$ & $125.0$ & $8.14$ & $11.7$ & $2.19 \ 10^5$ \\ 
    \hline
    $5$ & $10.14$ & $30.55$ & $7.88$ & $12.0$ & $2.19 \ 10^5$ \\
    \hline
    $6$ & $10.14$ & $12.77$ & $7.19$ & $12.8$ & $2.19 \ 10^5$ \\    
    \hline
    $7$ & $4.5$ & $125.0$ & $3.57$ & $11.1$ & $1.94 \ 10^5$ \\
    \hline
    $8$ & $4.5$ & $30.55$ & $3.46$ & $11.2$ & $1.94 \ 10^5$ \\
    \hline
    $9$ & $4.5$ & $12.77$ & $3.79$ & $12.0$ & $1.94 \ 10^5$ \\    
    \hline    
    \end{tabular}    
\end{center}
\caption{
Sea-state configurations selected from the \derisk{} database for illustration of the statistical model.
Each sea-state configuration considered in this study is identified by a number (first column).
The second column gives the nominal significant wave height imposed at the entry of the simulation domain, $\hsnom$.
The third column indicates the water depth at the considered station.
The fourth and fifth columns give, respectively, the effective significant wave height, $\hs$, and the mean wave period, $\tz$, measured at the considered station.
The last column reports the cumulative physical time of the simulations, calculated by summing the durations of all runs, with the initial $4500{\rm s}$ transient phase of each run omitted.
}    
\label{tab_Derisk_sims}
\end{table}

The parameter space covered by the \derisk{} simulations is shown in \fig{fig_Derisk_sims}, in the plane ($\depth / g{\tz}^2$, $\hs / g{\tz}^2$).
In this figure, each individual OceanWave3D simulation run is shown as a gray dot.
The nine sea states investigated in this study are highlighted as black dots in \fig{fig_Derisk_sims} and listed in \tab{tab_Derisk_sims}.
Each sea state appears as a cluster of black dots in \fig{fig_Derisk_sims}, because different simulations of the same sea state yield slightly different values of $\hs$ and $\tz$ over the finite simulation time.
Following the recommendations of Pierella et al. (2021) \cite{pierella_2021},
the first $4500 {\rm s}$ of each OceanWave3D run were discarded to remove the initial nonstationary phase during which the sea state gradually establishes in the simulation domain.
The values of $\hs$ and $\tz$ used in \fig{fig_Derisk_sims} were computed after discarding this initial nonstationary phase.
The values of $\hs$ and $\tz$ reported in \tab{tab_Derisk_sims} are averaged across the different runs of the corresponding sea-state configuration.

To demonstrate the model capabilities across the parameter space covered by the \derisk{} database, 
summary results are provided for all nine sea states in 
Section \ref{subsec_summary_results} (see \tab{tab_model_hyperparam}).
To limit the number of figures in this paper,
more detailed results are presented only for sea states \#1 and \#7 in Sections \ref{subsec_stoch_trajs}--
\ref{subsec_model_sensitivity}.
These two sea states were selected because they represent opposite extremes in terms of wave nonlinearities.
In the \derisk{} database, sea state \#7 is the most moderate, characterized by the smallest $\hsnom$ -- the significant wave height imposed at the entry of the simulation domain -- and the greatest water depth.
Conversely, sea state \#1 exhibits the largest value of the ratio $\hs / (g \tz^2)$, a nondimensional parameter that quantifies the degree of wave nonlinearity.
As will be illustrated below, the magnitude of wave nonlinearities has an important qualitative effect on how extreme stochastic trajectories behave relative to the rest of the population.

\subsection{Considered processes: selected kinematic variables and time window}
\label{subsec_kinVar}

A potential application of the statistical model is to estimate the distribution of slamming loads on a body exposed to wave impacts in a given sea state.
If the body is sufficiently small, it can be treated as a material point with respect to wave impact occurrence.
Wave impact events are then assumed to occur when the free surface upcrosses the material point, 
i.e., when the material point penetrates the water domain.
For illustrative purposes, the monitored material point is assumed to be fixed at the mean water level, at a given station.
A water-entry event is then defined as beginning when the free surface elevation $\eta$, measured at the station as a function of time $t$, upcrosses the mean water level, and ending when it reaches its first subsequent local maximum.
Hence, a water-entry event starts with $\de = \partial \eta / \partial t > 0$ and ends at the first subsequent zero crossing of $\de$.
To characterize a water-entry event, 
the following kinematic variables are considered:
\begin{itemize}
\item $\un$, the normal component of the fluid velocity at the free surface, defined as positive when pointing outward from the fluid domain.
\item $\sx = -\partial \eta / \partial x$, the slope of the free surface, measured in the direction of wave propagation (all sea states are unidirectional, with waves propagating in the direction of increasing $x$-coordinate). 
The slope $\sx$ is positive when the free surface is inclined downward in the direction of wave propagation.
\item $\azLag$, the vertical component of the fluid Lagrangian acceleration at the free surface, defined as positive when pointing upward.
\end{itemize}
The variables $\un$ and $\sx$ are selected for their relevance to the estimation of slamming loads (see next subsection for further details).
The third variable, $\azLag$, is introduced to account for the wave breaking limit in the analysis.
Note that the time-derivative of the free-surface elevation
is related to the variables $\un$ and $\sx$ through the free-surface kinematic condition:\footnote{
This form of the free-surface kinematic relation can be obtained starting from the usual form $\partial \eta / \partial t + u \partial \eta / \partial x = w$, where $u$ and $w$ are the horizontal and vertical components of the fluid velocity at the free surface, and using the identity $\cos[\atan (x)] = 1/\sqrt{1+x^2}$.
}
\begin{equation}
\label{eq_de}
\de = \un \sqrt{1+{\sx}^2}.
\end{equation}
Hence, $\de$ and $\eta$ can be deterministically reconstructed from $\un$ and $\sx$.

\subsection{Considered response variables}
\label{subsec_cons_resp_var}

In order to characterize the stochastic trajectories and test the model, the following response processes are introduced:
\begin{equation}
\left\{
\begin{aligned}
\Fx(t) & = \left( \frac{\tz}{\hs} \right)^2 \frac{\sx(t)}{\sqrt{1+{\sx(t)}^2}} {u_n(t)}^2 \label{eq_Fx} \\
\Fy(t) & = \left( \frac{\tz}{\hs} \right)^2 \frac{1}{\sqrt{1+{\sx(t)}^2}} {u_n(t)}^2 \, ,
\end{aligned}
\right.
\end{equation}
where the trigonometric identities $\sin[\atan (x)] = x/\sqrt{1+x^2}$ and  $\cos[\atan (x)] = 1/\sqrt{1+x^2}$
have been used. 
The factor $\left( \tz/\hs \right)^2$ has been introduced to nondimensionalize these two response processes.
The processes $\Fx$ and $\Fy$ can be regarded as proxies for the horizontal and vertical components of the slamming force resulting from a water-entry event.
Indeed, for a hydrodynamic impact occurring at constant velocity 
the slamming loads are proportional to the square of the water-entry velocity. 
The computation of the actual slamming force components would involve an additional time-varying slamming coefficient, which depends on the shape of the body, and an added-mass contribution proportional to the fluid acceleration.
The ability of the statistical model to capture the stochastic joint evolution of $(\un,\sx)$ over the water-entry phase
is tested using the following response variables:
\begin{align}
& \Fxmax = \max\limits_{t \in [0,\twe]} \Fx(t) \label{eq_kxmax} \\
& \Fymax = \max\limits_{t \in [0,\twe]} \Fy(t) \label{eq_kymax} \\  
& \Ix = \frac{1}{\tz}\int_0^{\twe} \Fx(t) \ \dd t \label{eq_ix} \\
& \Iy = \frac{1}{\tz}\int_0^{\twe} \Fy(t) \ \dd t \label{eq_iy} \, ,
\end{align}
where $t=0$ corresponds to the beginning of the water-entry event. 
The quantity $\Fxmax$ (resp. $\Fymax$) is the maximum value attained by $\Fx$ (resp. $\Fy$) over the water-entry duration, $\twe$,
and $\Ix$ (resp. $\Iy$) is the time integral of $\Fx$ (resp. $\Fy$) over the same period.
Similarly to $\Fx$ and $\Fy$, 
the variables $\Ix$ and $\Iy$ can be regarded as proxies for the horizontal and vertical components of the slamming impulse.
These four variables are particularly useful for testing the statistical model, as they combine the two processes $(\un,\sx)$ through nonlinear relations.
Moreover, $\Fxmax$ and $\Fymax$ are related to instantaneous values of $(\un,\sx)$ at times defined by the evolution of the system,
whereas $\Ix$ and $\Iy$ depend on the entire evolution of $(\un,\sx)$ over the water-entry duration.

\subsection{Dataset extraction}

To extract the dataset of water-entry events from the \derisk{} database, the following procedure was applied. For each simulation run, the first $4500$s of physical time were discarded (see \S\ref{subsec_select_case_studies}). 
The beginnings of water-entry events are identified by detecting zero upcrossings of the signal $\eta(t)$. The upcrossing definition implies that $\de$ and $\un$ are positive at upcrossings. 
The ends of water-entry events are identified by detecting the first subsequent downcrossing of zero by the signal $\de(t)$, which coincides with a zero downcrossing of $\un(t)$.
In the \derisk{} database, time series are sampled with a timestep of $0.14$s. For processing purposes, the time series are resampled onto finer grids using a modified Akima interpolation \cite{akima_1970, akima_1974, matlab_makima}. 
As an illustration,
the populations of water-entry trajectories extracted from the \derisk{} database for sea states $\#1$ and $\#7$ are shown in \fig{fig_wave_extraction}, with a selection of extreme trajectories highlighted in color (see the figure caption for details).

\def\scaleF{0.5}
\def\sizeS{-0.4cm}
\def\sizeSS{-0.4cm}

\begin{figure}[t!]
\vspace{-0.5cm}
\begin{center}
\begin{tabular}{cccc} 
	\hspace{\sizeSS{}}\centering \begin{turn}{90} \hspace{0.8cm} {\Large sea state \#1}\end{turn}  &
	\includegraphics[scale=\scaleF]{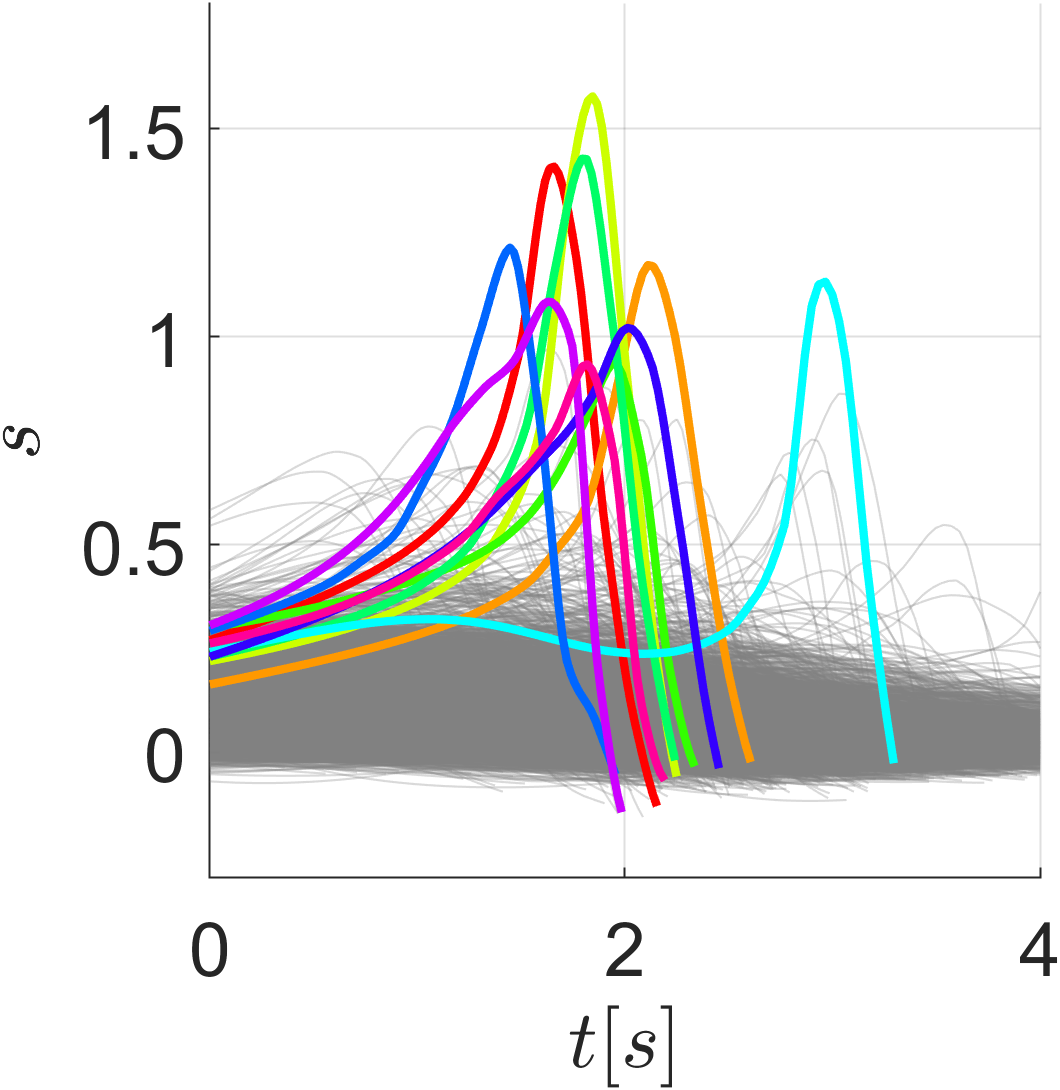} &
	\hspace{\sizeS{}}\includegraphics[scale=\scaleF]{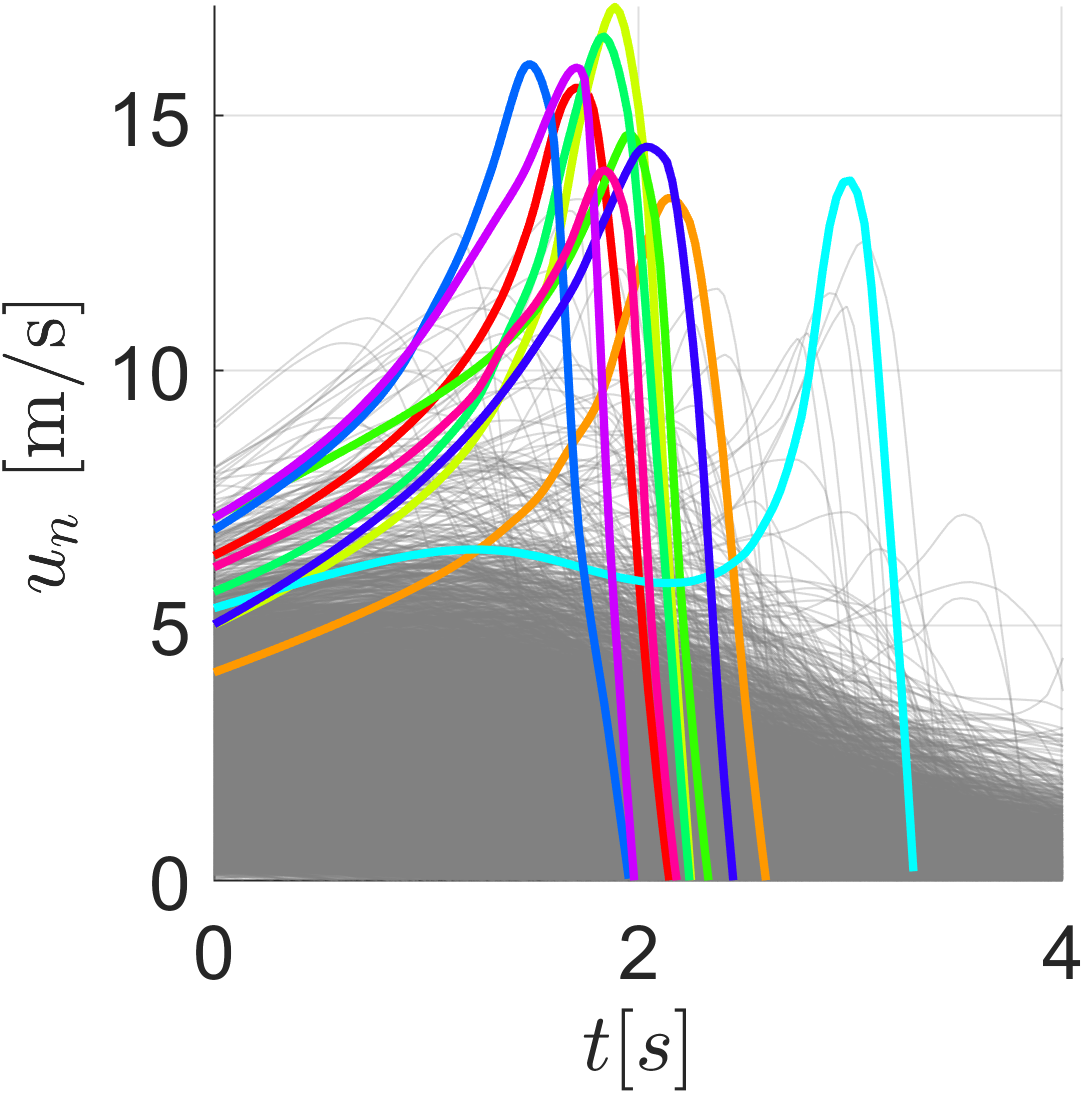} &
	\hspace{\sizeS{}}\includegraphics[scale=\scaleF]{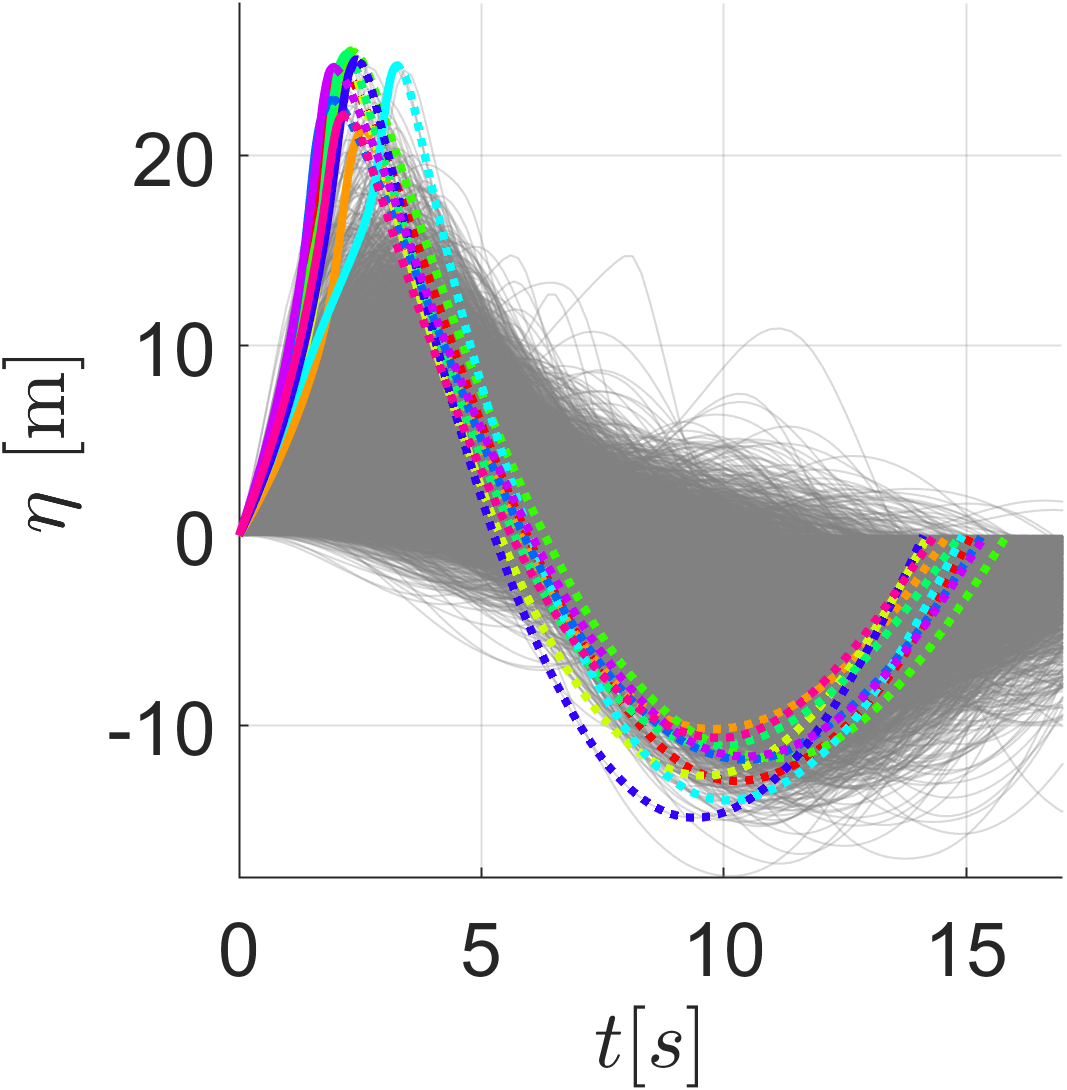}  \\
	\hspace{\sizeSS{}}\centering \begin{turn}{90} \hspace{0.8cm} {\Large sea state \#7}\end{turn}  &
	\includegraphics[scale=\scaleF]{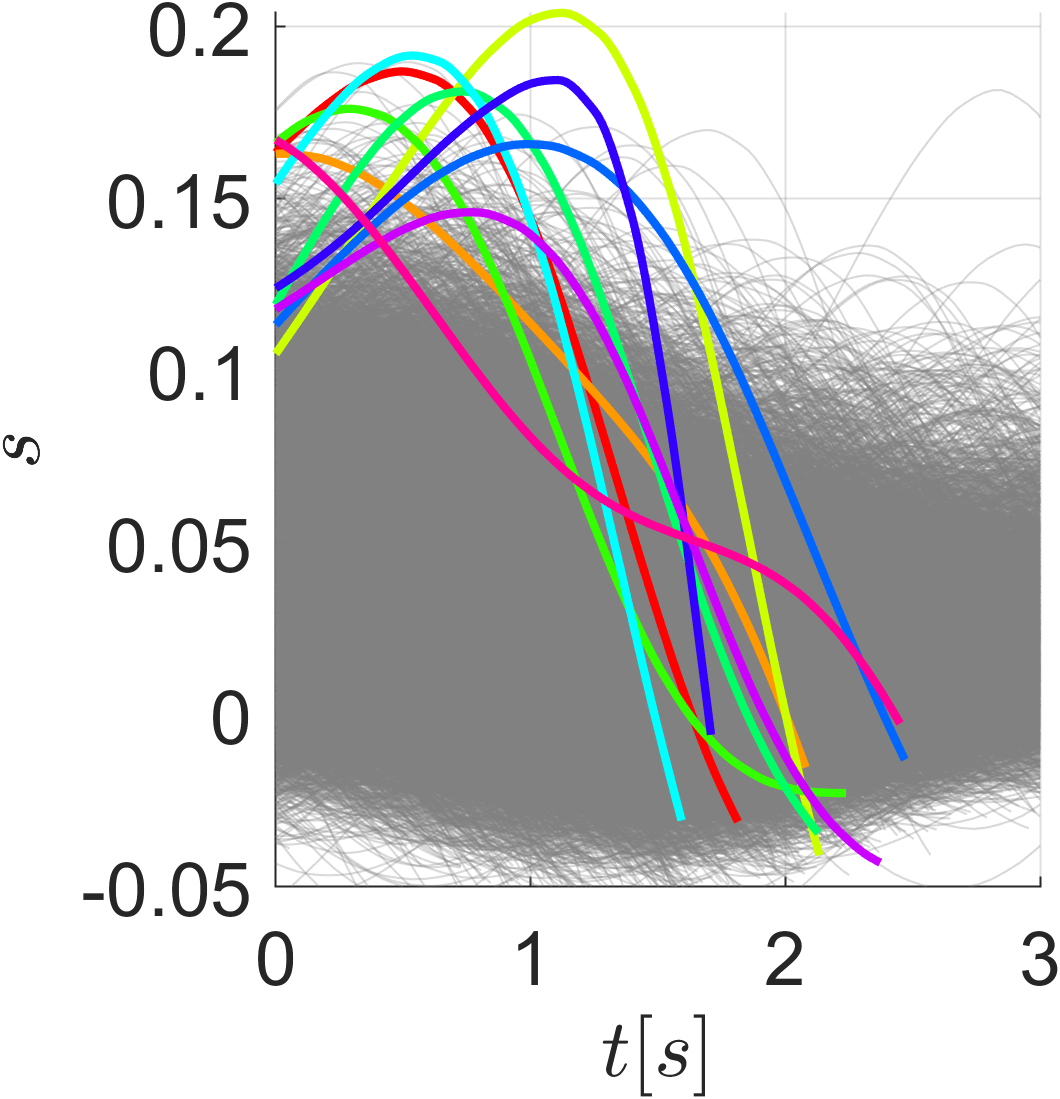} &
	\hspace{\sizeS{}}\includegraphics[scale=\scaleF]{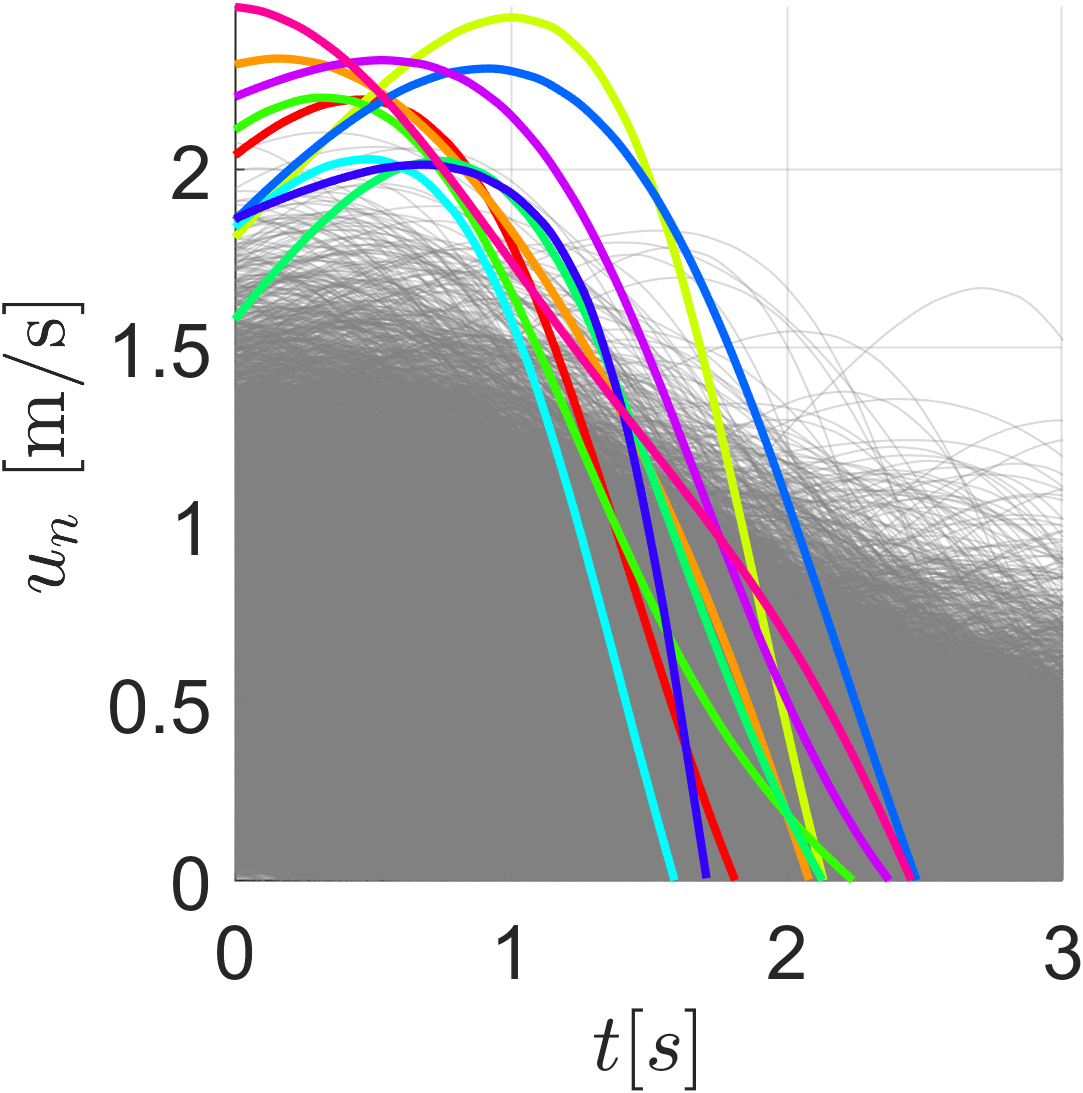} &
	\hspace{\sizeS{}}\includegraphics[scale=\scaleF]{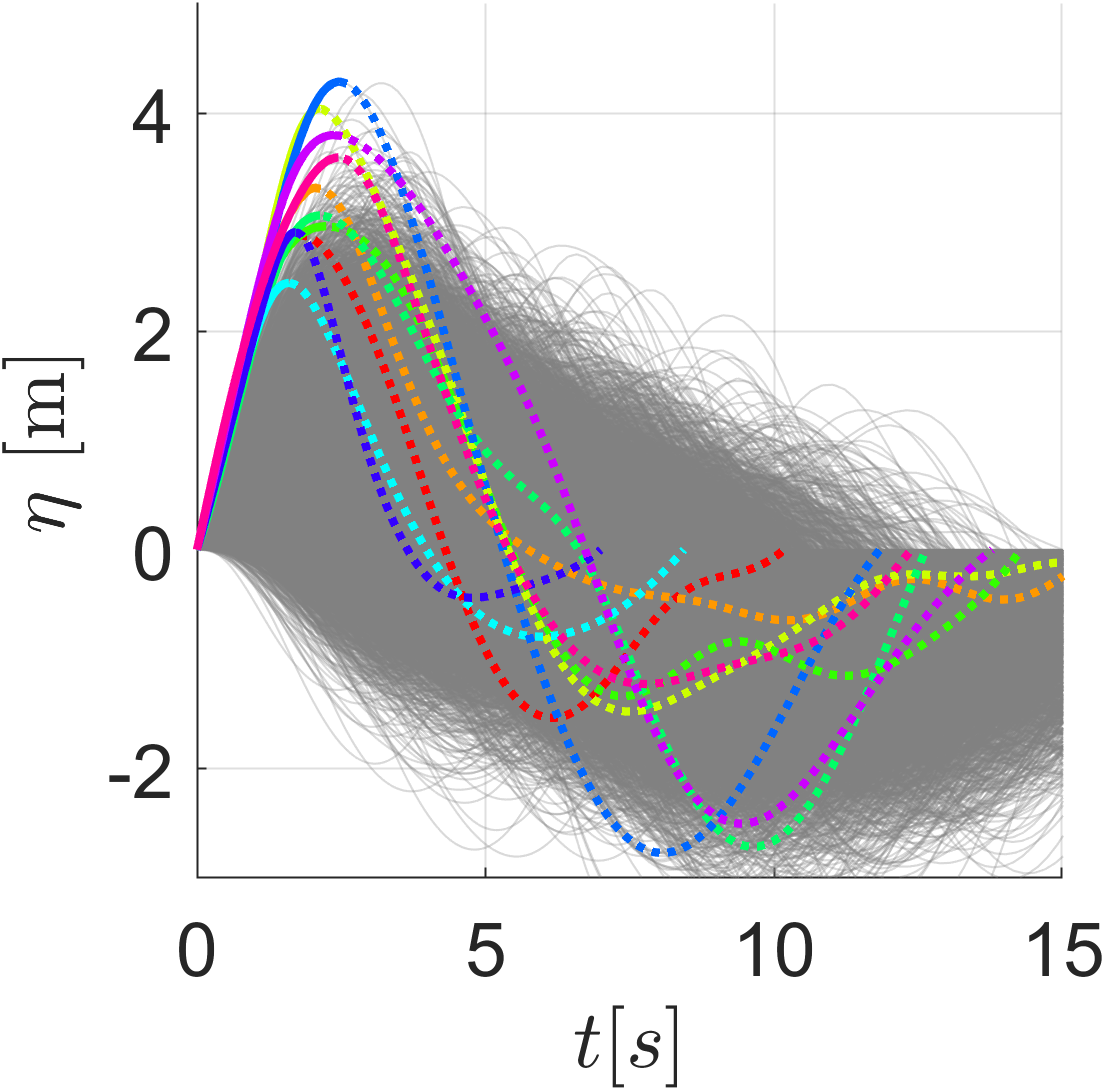} 	
\end{tabular}        
\end{center}
\caption{
Extraction of water-entry trajectories. The ``clouds'' of thin grey lines show the full dataset of stochastic trajectories extracted from the \derisk{} database. The thick colored lines highlight the $10$ most extreme trajectories in terms of $\Fxmax$ value.
Results are shown for sea states $\#1$ and $\#7$ in the first and second rows, respectively.
Trajectories of $s$ and $\un$ are shown in the first and second columns respectively. 
The corresponding wave profiles are shown in the third column. 
For the highlighted wave profiles, the solid portion (as opposed to the dotted portion) indicates the time interval corresponding to the water-entry phase. For a given water-entry event, the same color is used consistently across all three kinematic variables. 
}
\label{fig_wave_extraction}
\end{figure}

When fitting the model (see Sections \ref{sec_stocha_trajs_reduction}-\ref{sec_joint_distribution}), the water-entry trajectories extracted from a given \derisk{} simulation run are treated as statistically independent. 
Although this assumption is not strictly valid, the sequential autocorrelations observed in $\Fxmax$, $\Fymax$, $\Ix$, and $\Iy$ are already weak for non-successive events and become statistically non-significant for events separated by more than 10 waves.
In addition, it was checked that water-entry trajectories exhibiting extreme values of the response variables do not occur in clusters but are instead separated by a large number of waves.

\section{Dimensionality reduction: converting stochastic trajectories to feature vectors}
\label{sec_stocha_trajs_reduction}

This section presents the approach used to reduce a set of multivariate stochastic trajectories.

\subsection{Non-dimensionalizing the time parameter}
\label{subsec_adim_time}

As a preliminary step, the time parameter is nondimensionalized by the water-entry duration:
\begin{equation}
\label{eq_tadim}
\tadim = t/\twe \in [0,1] \, ,
\end{equation}
so that all stochastic trajectories are considered over a common nondimensional time interval.
To keep track of the water-entry duration, a natural choice would be to include $\twe$ in the feature vector.
As an alternative, the variable
\begin{equation}
\label{eq_Iun}
\Iun = \int_0^{\twe} \un(t) \ \dd t = \twe \int_0^{1} \un(\tadim) \ \dd \tadim  \, ,
\end{equation}
is considered to retain information about the water-entry duration. 
This alternative has been found to facilitate the fit of the bulk of the distribution.
This is because the dependence of $\Iun$ on the other features is less pronounced than that of $\twe$.

\subsection{Functional Principal Component Analysis} 
\label{subsection_fpca}

A Functional Principal Component Analysis (FPCA) 
is used to reduce the stochastic trajectories. 
FPCA consists in projecting the stochastic trajectories onto a set of orthogonal functions constructed so as to maximize the variance explained by the projections (see Chapter 8 of \cite{ramsay_2006} for a detailed introduction).
In the literature, these functions are referred to as "principal component curves", "eigenfunctions", or "empirical orthogonal functions". In the present paper, we use the term "principal component curve", or simply "principal component" for short.
The principal component curves are normalized, and the scalar products of a trajectory with these curves define its principal component scores.
The principal component scores of a given trajectory are collected into a feature vector, which serves as a reduced representation of the trajectory.
The dimension of the feature vector is 
\begin{equation}
\label{eq_ndim}
\nfeat = \npca + 1 \,
\end{equation}
where $\npca$ is the number of principal components retained in the analysis.
The additional feature corresponds 
to $\Iun$, which is used to keep track of the water-entry duration (see Eq. \ref{eq_Iun}).

Usually, only a limited number of principal component curves (and their associated scores) is sufficient to account for most of the variability observed among the stochastic trajectories.
In the present study, the stochastic trajectories of $\un$, $\sx$, and $\azLag$ -- all defined on a common nondimensional time grid (see Section \ref{subsec_adim_time}) -- must be modeled jointly. 
For simplicity, the principle of FPCA is presented below in a bivariate setting, namely for the joint modeling of $\un$ and $\sx$. 
Extending the approach to more than two variables is then straightforward.
FPCA can be performed either (i) separately on the variables $\un$ and $\sx$,
or (ii) jointly on both variables (see Section 8.5 in \cite{ramsay_2006}).
The first approach involves two separate sets of principal component curves, yielding two vectors of scores: one for $\un$ and one for $\sx$.
These two score vectors can then be concatenated into a single feature vector, together with the additional feature $\Iun$.
The second approach involves a set of composite principal component curves 
which yield a single vector of scores shared by $\un$ and $\sx$.
For the present study, the latter approach (joint FPCA) is preferred and adopted for several reasons:
\begin{itemize}
\item In the context of the present study, where the goal is to statistically model the stochastic trajectories of $\un$ and $\sx$, the number of principal components (and their associated scores) is treated as a hyperparameter.
When FPCA is performed separately on $\un$ and $\sx$, two hyperparameters must be specified.
In contrast, with a joint FPCA, only a single hyperparameter is required, making the model more parsimonious in terms of hyperparameters.
\item The stochastic trajectories of $\un$ and $\sx$ are strongly correlated, and their respective principal component curves -- when FPCA is performed separately on each variable -- exhibit similar shapes.
A joint FPCA naturally exploits this property and can explain the same amount of variance for $\un$ and $\sx$ with fewer principal component scores compared to separate FPCAs.
This renders the model more parsimonious by reducing the dimension of the feature vector.
\item As a corollary, a joint FPCA ensures that the resulting scores are uncorrelated, whereas a separate FPCA may yield correlations between the scores obtained for $\un$ and $\sx$. The absence of correlations among features is advantageous for the subsequent modeling of their joint distribution, as it removes linear dependencies between feature pairs.
\end{itemize}
The joint FPCA is sensitive to the relative scales of the quantities considered.
Here, $\un$ (a velocity) and $\sx$ (a slope) have different physical units. 
Therefore, as a preliminary step to the joint FPCA, 
the stochastic trajectories of $\un$ and $\sx$ are normalized 
by their respective time-averaged standard deviation:
\begin{equation}
\left\{
\begin{aligned}
& \sun  = \int_0^1 \sqrt{E[ \un(\tadim)^2] - E[ \un(\tadim)]^2} \ \dd \tadim \\
& \ssx  = \int_0^1 \sqrt{E[ \sx(\tadim)^2] - E[ \sx(\tadim)]^2} \ \dd \tadim \, ,
\end{aligned}
\right.
\end{equation}
where $E$ denotes the expectation operator.
This ensures that $\un$ and $\sx$ contribute comparably to the construction of the composite principal components.
Let the $i^{\rm th}$ principal component be denoted by 
\begin{equation*}
P_{i} = [{P_i}^{(\un)}(\tadim), {P_i}^{(\sx)}(\tadim)]_{\tadim \in [0,1]}
\end{equation*}
where ${P_i}^{(\un)}$ (resp. ${P_i}^{(\sx)}$) is the $\un$-curve (resp. $\sx$-curve) of the composite principal component.
The principal components are orthogonal by construction, and normalized for convenience. In the case of composite principal components, these two conditions read:
\begin{equation}
\int_0^{1} \left[ {P_i}^{(\un)}(\tadim) {P_j}^{(\un)}(\tadim) 
+ {P_i}^{(\sx)}(\tadim) {P_j}^{(\sx)}(\tadim)
\right] {\rm d} \tadim =
\left\{
\begin{aligned}
 & 1 \,, \ \ \ \ \ \text{if $i=j$} \\
 & 0 \,, \ \ \ \ \ \text{if $i\ne j$}
\end{aligned}
\right.
\end{equation}
With regard to numerical implementation, the FPCA was performed using the \texttt{fda} package \cite{fda_package} in a R environment.

\subsection{Reconstruction of trajectories} 
\label{subsec_reconstruct_trajectories}

Based on $\npca$ principal components, the stochastic trajectories can be reconstructed as follows:
\begin{equation}
\label{eq_pca_reconstruction}
\left\{
\begin{aligned}
\widehat{\un} (\tadim) & = \sun \sum_{i=1}^{\npca} x_i \cdot {P_i}^{(\un)}(\tadim) \\
\widehat{\sx} (\tadim) & = \ssx \sum_{i=1}^{\npca} x_i \cdot {P_i}^{(\sx)}(\tadim) \, ,
\end{aligned}
\right.
\end{equation}
where $(x_i)_{i=1:\npca}$ are the principal component scores. The scales $\sun$ and $\ssx$ account for the fact that the FPCA was performed on normalized data.
The timescale of the reconstructed trajectories is then recovered by remembering that 
the last feature stores the value of $\Iun$:
\begin{equation}
\label{eq_twe_rec}
\widehat{\twe} =  \frac{\Iun}{\int_0^1 \widehat{\un}(\tadim) \dd \tadim } \, .
\end{equation}
The reconstruction errors, $(\widehat{\un}(\tadim)-\un(\tadim))_{\tadim \in [0,1]}$, $(\widehat{\sx}(\tadim)-\sx(\tadim))_{\tadim \in [0,1]}$, and $\widehat{\twe}-\twe$, become smaller as the number of principal components retained, $\npca$, increases. 

\section{Modeling the distribution of the feature vector}
\label{sec_joint_distribution}

This section describes the statistical modeling of the multivariate distribution underlying the feature vector.
The feature vector $\Xv = [X_1, X_2, \dots, X_d]$ consists of the PCA scores, $X_1 = x_1, \dots, X_{d-1} = x_{d-1}$, together with the additional variable $X_d = \Iun$ (see Eq. \ref{eq_Iun}).
Thus, the dimension of the problem is $d = \npca + 1$, where $\npca$ denotes the number of retained principal components.
Both the bulk and the tail of the multivariate distribution are modeled, with the marginal distributions and the dependence structure treated separately.
The modeling of the marginal distributions is described in \S\ref{subsec_fit_margins}.
The multivariate structure is addressed in \S\ref{subsubsec_bulk_model} for the bulk of the distribution, and in \S\ref{subsec_HT_model} for the tail.

\subsection{Fitting the marginal distributions of the feature vector $\Xv$}
\label{subsec_fit_margins}

The distribution of each margin $X_i$, $i \in \{ 1, \cdots, d \}$, is modeled using a composite approach. 
The bulk of the distribution is modeled using kernel density estimation, while the upper tail is modeled with a generalized Pareto distribution (GPD) and the lower tail with a reversed GPD.
The number of data points $\nMT$ 
(the subscript "MT" refers to marginal tail)
used to estimate the parameters of each GPD model is treated as a hyperparameter.
The upper-tail GPD (resp. lower-tail reversed GPD) is connected to the kernel density estimate at a threshold $u_i$ (resp. $v_i$).
The thresholds are defined as 
\begin{align}
u_{i} = & 0.5 [{x_i}^{(\nobs-\nMT)}+{x_i}^{(\nobs-\nMT+1)}] \, , \text{ for the upper tail} \\
v_{i} = & 0.5 [{x_i}^{(\nMT)}+{x_i}^{(\nMT+1)}] \, , \text{ for the lower tail}
\end{align}
where ${x_i}^{(k)}$ denotes the $k^{\rm th}$ smallest observed value of $X_i$ in the dataset.
The resulting composite model for the cumulative distribution function of $X_i$ is:
\begin{equation}
\label{eq_Fi}
    F_i(x)= 
\begin{dcases}
     \tilde{F}_{i}(v_{i}) 
     \SGPD\left( -\frac{x-v_{i}}{{s_i}^{(l)}} ; {\xi_i}^{(l)} \right) 
     & \text{for } x < v_{i} \\
     \tilde{F}_{i}(x)   & \text{for } u_{i} \ge x \ge v_{i} \\
     1 - [1 - \tilde{F}_{i}(u_{i}) ] \SGPD\left( \frac{x-u_{i}}{{s_i}^{(u)}} ; {\xi_i}^{(u)} \right) & \text{for } x > u_{i}    \, .
\end{dcases}
\end{equation}
Here, $\SGPD(.;\xi)$ denotes the complementary cumulative distribution function of the standard GPD: 
\begin{equation}
\SGPD(z;\xi) = 
\max(\{ 1 + \xi z \},0)^{-1/\xi} \, , \text{ for } z\ge 0 \text{ and } \xi \in (-\infty,\infty).
\end{equation}
The shape and scale parameters of the upper (resp. lower) tail, ${\xi_i}^{(u)}$ and ${s_i}^{(u)}$ (resp. ${\xi_i}^{(l)}$ and ${s_i}^{(l)}$), are estimated using
the $L$-moment approach \cite{hosking_1990, hosking_1997}.
For the numerical implementation, the Matlab library developed by G. Talbot was employed \cite{evsa_lib}.
Compared with maximum likelihood estimation, the $L$-moment method provides estimates with greater stability and robustness against model misspecification, data errors, or outliers, thereby reducing the sensitivity of the estimates to the selection of thresholds $u_i$ and $v_i$.

\subsection{Bulk of the distribution: simplified vine copula model}
\label{subsubsec_bulk_model}

The bulk of the distribution of the feature vector $\Xv$ is modeled using a simplified vine copula approach.
The principles underlying this approach are briefly described in \ref{appendix_vine_copula}.
For numerical implementation, the R package \texttt{rvinecopulib} developed by Nagler and Vatter \cite{rvinecopulib_package} was used.
This package provides several options to estimate a copula distribution based on a simplified vine copula representation.
In the present study, the following options were selected:
\begin{itemize}
\item The structure of the regular vine, representing the decomposition as illustrated in \eq{eq_cop_decomp_simp},
is selected using a criterion based on Kendall's $\tau$ estimates.
\item The densities of the bivariate copulas are estimated nonparametrically using a local likelihood estimator of order 2 \cite{loader_1999, geenens_2017, nagler_2017}.
\end{itemize}

\subsection{Multivariate tail: conditional approach}
\label{subsec_HT_model}

\subsubsection{Partition of the tail domain into subsets}

The multivariate tail of the feature vector $\Xv$ is modeled using 
the conditional approach of Heffernan and Tawn (2004) \cite{heffernan_2004},
together with the subsequent improvements proposed in \cite{keef_2013}.
As a preliminary step,
the vector $\Xv$ is transformed componentwise so that
the margins of the resulting vector $\Yv$ follow the standard Laplace distribution:
\begin{equation}
Y_i = {\FLap}^{-1}  (F_i(X_i)) \, \text{, for } i \in \{ 1, \cdots , d \} \, ,
\end{equation}
where $F_i$ is defined in \eq{eq_Fi}.
The cumulative distribution function of the standard Laplace distribution is: 
\begin{equation}
    \FLap(x)= 
\begin{dcases}
     \frac{1}{2} \exp(x) ,& \text{if } x\leq 0\\
    1 - \frac{1}{2} \exp(-x),              & \text{if } x> 0 \, .
\end{dcases}
\end{equation}
A threshold $\yth$, above which the approach is applied, must be chosen.
This threshold corresponds to an exceedance probability $\eHT = 1-\FLap(\yth) = \FLap(-\yth)$, where the subscript "HT" refers to Heffernan \& Tawn.
In Section \ref{sec_restults}, $\eHT$ is parametrized as 
\begin{equation}
\label{eq_eps_HT}
\eHT = \nHT /\nobs \, ,
\end{equation}
where $\nobs$ is the total number of data points in the dataset, and $\nHT$ is a hyperparameter that can be interpreted as the nominal number of data points used to fit the HT model.\footnote{
The actual number of data points above the threshold $\yth$ may differ slightly from $\nHT$. 
}
The multivariate distribution tail of $\Yv$ is then modeled by considering separately the regions:
\begin{equation}
\mathcal{S}_i = \{ \Yv \in \mathbb{R}^d : \sgn(i) Y_{\abs{i}} > \yth  \} \, \text{, with } i \in \{ -d, \cdots,-1,1, \cdots d \} .
\end {equation}
The index $i$ is allowed to take negative values to incorporate lower-tail regions.
The corresponding regions in the original $\Xv$-space can be written as:
\begin{align}
\mathcal{R}_{i}  = \{ \Xv \in \mathbb{R}^d : \sgn(i) \CDFXi(X_i) > - \eHT + 0.5(1+\sgn(i))  \}  \, \text{, with } i \in \{ -d, \cdots,-1,1, \cdots d \} \, .
\end {align}

\subsubsection{Conditional model in a given subset}
\label{subsubsec_ht_condi_model}

To simplify notation in this section, the Heffernan \& Tawn conditional approach is described 
for a region $\mathcal{S}_{i}$ with $i>0$. 
The extension to a region $\mathcal{S}_{i}$ with $i<0$ simply requires replacing $Y_i$ by $-Y_i$.
Let $\Yri$ denote the vector $\Yv$ with its $i^{\text{th}}$ component removed.
The conditional approach models the random vector $\Yri$, given $\Yi$, as:
\begin{equation}
\label{eq_ht_reg_form}
\Yri  = \ai(\Yi) + \bi(\Yi) \Zi  
\end{equation}
where $\ai : [\yth,+\infty) \rightarrow \mathbb{R}^{d-1}$ 
and $\bi : [\yth,+\infty) \rightarrow \mathbb{R}^{d-1}$ 
are, respectively, location and scale functions, and where $\Zi \in \mathbb{R}^{d-1}$ is a residual random vector independent of $\Yi$.
For a wide class of distributions for $\Xv$, 
Keef et al. (2013)
\cite{keef_2013} showed that the location and scale functions take the asymptotic forms:
\begin{equation}
\label{eq_form_loc_scale_functions}
\begin{dcases}
\ai (t) & = \aai t \\
\bi (t) & = t^{\bbi} 
\end{dcases}
\end{equation}
where $t \in [\yth,+\infty)$, and $(\aai, \bbi)\in[-1,1]^{d-1}\times(-\infty,1)^{d-1}$ are constant vectors.
The exponentiation $t^{\bbi}$ is understood componentwise.
Eqs. (\ref{eq_ht_reg_form}-\ref{eq_form_loc_scale_functions}) 
constitute the model structure used to describe the multivariate tail of an unknown distribution.

The distribution of the residual vector $\Zi$ 
may take a wide range of forms depending on the distribution of $\Xv$.
Heffernan and Tawn proposed a semi-parametric strategy to model $\Zi$.
The first step falsely assumes that the residual takes the form $\Zi = \mi + \si \textbf{N}$,
where $\textbf{N}$ is a standard normal vector in $\mathbb{R}^{d-1}$,
and where $\mi$ and $\si$ denote the mean and scale vectors of $\Zi$.
Under this working assumption, the vectors $(\aai, \bbi, \mi, \si)$ are estimated jointly using maximum likelihood estimation (MLE).
Using $\widehat{\aai}$ and $\widehat{\bbi}$, the MLE estimates of $\aai$ and $\bbi$,
the second step then estimates the distribution of 
\begin{equation}
\widehat{\Zi} = \frac{\Yri - \widehat{\aai} \Yi}
{\Yi^{\widehat{\bbi}}} \, ,
\end{equation}
where exponentiation and division are understood componentwise.
Estimating this distribution is challenging because the number of observations used 
to fit a Heffernan \& Tawn model is typically on the order of a hundred, while $\Zi$ may be high-dimensional.
Following the suggestion of Keef et al. (2013) \cite{keef_2013}, the distribution of $\widehat{\Zi}$ is approximated by the empirical distribution of the observed sample. 
Note that this does not induce as severe a sparsity in simulated data as one might expect, since $\Zi$ appears only as a residual in \eq{eq_ht_reg_form}, while $\Yi$ is sampled continuously.

\subsubsection{Selecting and assembling the conditional models}
\label{subsubsec_sel_assemb_HT}

Hereafter, the conditional HT model obtained for the region $\mathcal{R}_i$
is formally denoted by $\mHT$.
In the most general setting, modeling the full multivariate tail -- including the ranges of extremely small values --
would require fitting $2d$ conditional HT models, $\mHT$, 
with $i={-d, \cdots, -1,1,\cdots, d}$.
In practice, the model can be made more parsimonious
by restricting the set of HT conditional models considered.
Below, the set of indices:
\begin{equation}
\seqHT = \{ i_{1}, \cdots i_{p} \} \, , \text{ with $\seqHT \subset \{ -d, \cdots, -1,1,\cdots, d \}$} \, ,
\end{equation}
specifies which HT conditional models are actually considered.
Then, the $\Xv$-domain covered by the HT conditional models is: 
\begin{equation}
\label{eq_HT_regions}
\rHTall =  \bigcup\limits_{ i \in \seqHT} \rHT   \, .
\end{equation}
The intersection of two distinct extreme subsets, $\mathcal{R}_i$ and $\mathcal{R}_j$, has nonzero volume if $i \ne -j$.
To prevent the HT conditional models from covering common regions,
the domain assigned to each model in the $\Yv$-space are restricted as follows:
\begin{equation}
\label{eq_constraint_no_overlap}
   {\mathcal{S}_i}^{(r)}  = \mathcal{S}_i \cap \mathcal{U}_i   \, ,
 \end{equation}
with
\begin{equation}
\label{eq_region_restriction}
 \mathcal{U}_i =  \{ \Yv \in \mathbb{R}^d : \sgn(j) Y_\abs{j} \le \sgn(i) Y_\abs{i} \, , \ j \in \seqHT \, , \abs{j} \ne \abs{i}   \}  \, .
\end{equation}
The set $\mathcal{U}_i$ corresponds to the region where $Y_\abs{i}$ is the "most extreme" among the components of $\Yv$ that are modeled by a HT conditional model.

\begin{algorithm}[b!]
\caption{Simulating a realization of $\Xv$ from the statistical model}
\label{algo_simu}
\LinesNumbered
 Draw a region from the set $\{ \rBulk, \rHTall \}$.
The probability of selecting $\rHTall$ is $\pHT = \nobsHT / \nobs$,
with $\nobs$ the total number of data points in the training dataset
and $\nobsHT$ the number of those contained in the region $\rHTall$. \label{line:pht}\\
accepted $\leftarrow$ False \\
\If{$\rHTall$ is drawn}{
  \While{not accepted}{
  Draw one HT model, $\mHTk$, from the list $(\mHT)_{i \in \seqHT}$. 
Each HT model is selected with equal probability. \label{line:drawMk} \\
  Generate $Y_k=\sgn(k)(A+\yth)$, where $A$ is drawn from the standard exponential distribution. \label{line:Lap}\\
  Generate $\Zk$ from the empirical distribution of $\widehat{\Zk}$.\\
  Compute $\Yrk$ from \eq{eq_ht_reg_form}.\\
  $\Yv \leftarrow$  concatenate$(Y_k, \Yrk)$ \\
  \If{$\Yv \in {\mathcal{S}_k}^{(r)}$}{
    accepted $\leftarrow$ True
  }  
  }
\For{$i \gets 1$ \KwTo $d$}{
$X_i \leftarrow {\CDFXi}^{-1}(\FLap(\Yi))$ \\
}
$\Xv \leftarrow$ concatenate$(X_1, \cdots, X_d )$\\
}
\Else{
  \While{not accepted}{
  Generate $\Xv \sim \mBulk$ \\  
  \If{$\Xv \in \rBulk$}{
    accepted $\leftarrow$ True
  }
  }
}
\Return{$\Xv$}
\end{algorithm}

\subsection{Simulating the model}
\label{sec_simu_model}

The procedure for simulating the global model is described in this subsection.
The region of the $\Xv$-space covered by the collection of HT conditional models is denoted $\rHTall$ (see Eq. \ref{eq_HT_regions}).
The complement of this region,
\begin{equation}
\rBulk' =  {\rHTall}^\complement  \,  ,
\end{equation}
is modeled using $\mBulk$, the simplified-vine-copula-based model that represents the bulk of the distribution.
Since the vine copula model is non-parametric, it does not accurately capture the dependency structure in the multivariate tail.
Therefore the region allocated to the vine copula model is further restricted to 
\begin{equation}
\rBulk =  \rBulk' \cap \mathcal{I}  \,  ,
\end{equation}
where 
\begin{equation}
\mathcal{I} = \bigcap\limits_{ i \in 0:d} I_i \, , \ \ \ \ I_i = \{ \Xv \in \mathbb{R}^d : {x_i}^{(n)} > X_i >   {x_i}^{(1)} \} \, ,
\end{equation}
and ${x_i}^{(1)}$ and ${x_i}^{(n)}$ are the smallest and largest values of $X_i$ observed in the dataset used to fit the model.
Based on this subdivision of the $\Xv$-space, 
a synthetic realization of $\Xv$ is drawn using 
Algorithm \ref{algo_simu}. Some comments about the algorithm follow:
\begin{itemize}
    \item The algorithm is based on an acceptance-rejection scheme.
    \item In line \ref{line:pht}, the probability of drawing the synthetic realization from the region  $\rHTall$ is computed as $\pHT = \nobsHT / \nobs$. 
    Note that $\nobsHT$ should not be confused with $\nHT$, the latter being the hyperparameter introduced in \eq{eq_eps_HT}.  
    \item In line \ref{line:drawMk}, each HT model is selected with equal probability, since the same threshold $\yth$ is used for all HT conditional models. If different thresholds are employed, the probability weights should be adjusted accordingly.
    \item Line \ref{line:Lap} generates $Y_k$ from a standard Laplace distribution, conditioned on $\sgn(k) Y_k>\yth$.
\end{itemize}

\subsection{Enforcing the wave breaking limit in the stochastic model}
\label{sec_enforce_breaking_limit}

The modeling approach presented above is purely data-driven and does not explicitly enforce physical constraints such as the wave-breaking limit. 
The statistical model is trained on the \derisk{} dataset, which incorporates a wave-breaking filter, so the model may implicitly capture the wave breaking limit in a data-driven manner. 
However, there is no guarantee that the synthetic waves generated by the model respect this physical constraint.

One way to address this issue is to explicitly enforce the wave-breaking limit using a wave-breaking criterion.  
Wave-breaking criteria are typically classified into three categories: geometric, kinematic, and dynamic (see, e.g., \cite{wu_2002, babanin_2011}).
In this study, we adopt a dynamic criterion similar to that used for the breaking filter in the \derisk{} simulations (see condition (\ref{eq_breaking_filter})):
\begin{equation}
\label{eq_breaking_filter_my}
-\azLag < \hat{\beta} g \, , 
\end{equation}
with $\hat{\beta}=0.6$.
To enforce this limit in the stochastic model, synthetic trajectories that violate condition (\ref{eq_breaking_filter_my}) are rejected.
Naturally, enforcing this criterion requires that the kinematic variable $\azLag$ be included in the stochastic model.
The coefficient $\hat{\beta}$ has been set slightly above $0.5$ to remain consistent with the \derisk{} dataset, where vertical Lagrangian accelerations as low as $\simeq - 0.6 g$ were observed. 
This moderate overshoot of $-0.5 g$ in the \derisk{} simulations is likely due to the fact that the breaking filter acts as a smoothing filter rather than as a sharp cutoff (see \cite{pierella_2021} for details).

\section{Strategy for hyperparameter selection}
\label{seq_strategy_hyperparams}

In the implementation of the stochastic model described in Section \ref{sec_joint_distribution}, 
the number of hyperparameters was reduced to three:
\begin{itemize}
\item $\npca$: the number of components considered in the principal component analysis.
\item $\seqHT$: the set of conditional HT models enabled.
\item $\nT$: the number of observations used for tail modeling. To limit the overall number of hyperparameters, the number of observations used to fit each marginal tail model and each conditional HT model are set equal, i.e., $\nMT = \nHT = \nT$.
\end{itemize}
For each sea state configuration, the hyperparameters were chosen in a stepwise manner, following the strategy presented below. 

\subsection{Step 1: choosing the number of principal components}
\label{subsec_step1_npca}

As a first step, the number of principal components, $\npca$, 
is chosen such that the reconstruction error mentioned in Section \ref{subsec_reconstruct_trajectories} is below some threshold.  
In principal component analysis, the ratio of explained variance is commonly used as a metric to quantify the reconstruction error averaged over the considered dataset.
In the present study, extreme events are considered, so the explained variance may not be a relevant metric. 
Instead, a metric based on the empirical quantiles of the response variables is defined as follows:
\begin{equation}
\label{eq_score_pca} 
{r_v}^{\rm (pca)} = \sqrt{ \mathlarger{\sum}_{k=n-\mTail+1}^{n} \left( \frac{\widehat{q_v}^{(k)}}{{\overline{q}_v}^{(k)}} - 1 \right)^2} \, ,
\end{equation}
where $v$ denotes the considered response variable -- i.e., $\Fxmax$, $\Fymax$, $\Ix$, or $\Iy$ -- and $n$ is the size of the dataset.
The terms ${\overline{q}_v}^{(k)}$ and $\widehat{q_v}^{(k)}$ are the $k^{\rm th}$ smallest values obtained for the response variable $v$, from the true trajectories and PCA-reconstructed trajectories, respectively.
The number of considered quantiles, $\mTail$, is a parameter that must be chosen a priori: the smaller $\mTail$, the more focused on extreme values the score $r^{\rm (pca)}_v$ becomes.
Then, for a fixed value of $\mTail$, the number of principal components, $\npca$, is selected such that the score $r^{\rm (pca)}_v$ falls below a prescribed threshold for each of the four response variables considered.

\subsection{Step 2: choosing the HT conditional models}
\label{subsec_strategy_step2}

Once the number of principal components has been chosen, 
the next step consists in the selection of the enabled HT models, identified by the set of indices $\seqHT$ (see Section \ref{subsubsec_sel_assemb_HT}).
The set $\seqHT$ is chosen according to the following criteria.
(i) The set $\seqHT$ should be chosen so that all trajectories considered as ``extreme''
fall within the region $\rHTall$ defined in \eq{eq_HT_regions}.
This ensures that all extreme trajectories contribute to the modeling of the multivariate tail.
Here, extreme trajectories are defined as those producing extreme values of $\Fxmax$. Another response variable could have been used, since the response variables introduced in Section \ref{subsec_cons_resp_var} tend to be jointly extreme.
(ii) It is desirable to keep the model as parsimonious as possible. Combining these two objectives, the set $\seqHT$ is chosen to have the smallest possible cardinality while still satisfying constraint (i).

\subsection{Step 3: choosing the threshold for tail modeling}
\label{subsec_step3_nt}

As a third step, the number of observations $\nT$ used for tail modeling is selected.
To evaluate the quality of the model obtained for a given value of $\nT$, a score similar to \eq{eq_score_pca} is introduced:
\begin{equation}
\label{eq_score_mod}
{r_v}^{\rm (mod)} = \sqrt{ \mathlarger{\sum}_{k=n-\mTail+1}^{n} \left( \frac{{q_v}^{(k)}}{{\overline{q}_v}^{(k)}} - 1 \right)^2} \, ,
\end{equation}
where ${q_v}^{(k)}$ is the $((k-1/2)/n)$-quantile predicted by the stochastic model for the response variable $v$.
The hyperparameter $\nT$ is constrained to lie in an admissible range.
Then, in this range, the hyperparameter $\nT$ is chosen such that the score ${r_v}^{\rm (mod)}$ is minimized for a given response variable.
Below, the response variable $\Fxmax$ is used for this task, as it was found to be the most challenging response variable for the model.

\section{Results}
\label{sec_restults}

To illustrate the proposed approach, 
Section \ref{subsec_summary_results} 
presents the results obtained 
for the nine sea states listed in \tab{tab_Derisk_sims}.
More detailed results are then provided in Sections 
\ref{subsec_stoch_trajs}--\ref{subsec_model_sensitivity} 
for sea states $\#1$ and $\#7$.

\begin{table}[b!]
\begin{center}
    \begin{tabular}{| c | c || c | c | c || c | c | c  |}
    \hline
    \multicolumn{2}{|c||}{Dataset} & \multicolumn{3}{c||}{Hyperparameters} & \multicolumn{3}{c|}{Results} \\
    \hline
    $\#$ sea state  & $\nobs$  & $\npca$ & $\nT$  & $\seqHT$ & $r^{\rm (pca)}_{\Fxmax}$ & $r^{\rm (mod)}_{\Fxmax}$ & $\dquant$ \\ 
    \hline
    1 & $7679$ & $31$ & $50$ & $\left\{1, d \right\}$ & $0.022$ & $0.20$ & $0.23$ \\
    \hline
    2 & $7386$ & $30$ & $200$ & $\left\{1, d \right\}$ & $0.017$ & $0.05$ & $0.05$ \\
    \hline
    3 & $7335$ & $27$ & $200$ & $\left\{1, d \right\}$ & $0.030$ & $0.04$ & $0.02$ \\
    \hline
    4 & $18699$ & $32$ & $200$ & $\left\{1, d \right\}$ & $0.030$ & $0.14$ & $0.09$ \\
    \hline
    5 & $18246$ & $27$ & $200$ & $\left\{1, d \right\}$ & $0.024$ & $0.08$ & $0.06$ \\
    \hline
    6 & $17023$ & $29$ & $125$ & $\left\{1, d \right\}$ & $0.030$ & $0.01$ & $0.01$ \\
    \hline
    7 & $17537$ & $14$ & $125$ & $\left\{1, d \right\}$ & $0.027$ & $0.04$ & $0.00$ \\
    \hline
    8 & $17422$ & $13$ & $125$ & $\left\{1, d \right\}$ & $0.014$ & $0.09$ & $0.01$ \\
    \hline
    9 & $16184$ & $15$ & $50$ & $\left\{1, d \right\}$ & $0.026$ & $0.11$ & $0.00$ \\
    \hline
    \end{tabular}    
\end{center}
\caption{
Configurations considered in Section \ref{sec_restults}.
The first two columns relate to the dataset under consideration:
the first column indicates the sea state, and
$\nobs$ denotes the dataset size, i.e., 
the number of stochastic trajectories extracted from the \derisk{} database.
The next three columns specify the hyperparameters of the stochastic model:
$\npca$ is the number of components retained in the functional PCA; $\nT$ is the number of observations used to fit each marginal tail and each HT conditional model;
$\seqHT$ indicates, following the convention described in Section \ref{subsubsec_sel_assemb_HT}, which HT conditional models are enabled.
The dimension $d$ is related to $\npca$ through \eq{eq_ndim}. 
The last three columns report the values of the three metrics, $r^{\rm (pca)}_{\Fxmax}$, $r^{\rm (mod)}_{\Fxmax}$, and $\dquant$, defined in Eqs. 
(\ref{eq_score_pca}--\ref{eq_score_extreme_quant}), 
respectively.
The metric $r^{\rm (mod)}_{\Fxmax}$ is computed with the breaking filter enabled (see Section \ref{sec_enforce_breaking_limit}).
}    
\label{tab_model_hyperparam}
\end{table}

\subsection{Model hyperparameters and summary results}
\label{subsec_summary_results}

The model hyperparameters are summarized in \tab{tab_model_hyperparam}.
The dimension of the stochastic model is $\nfeat = \npca+1$, where $\npca$ denotes the number of principal components used to represent the multivariate trajectories (see Section \ref{subsection_fpca}).
Following the strategy described in Section \ref{subsec_step1_npca}, $\npca$ was chosen as being the smallest number of principal components yielding a score $r^{\rm (pca)}_v$ below $0.03$ with $\mTail=20$, for each of the four response variables, $\Fxmax$, $\Fymax$, $\Ix$, and $\Iy$.
The required set of conditional HT models was found to be parsimonious for all sea states considered. 
Indeed, the strategy described in Section \ref{subsec_strategy_step2}
indicated that $\seqHT = \{ 1, d \} $ provides an appropriate selection of HT models for all considered sea states.
The indices $1$ and $d$ correspond to the HT models in which the conditioning variables are the score of the first principal component and $\Iun$, respectively.
Finally, the third hyperparameter was determined following the strategy described in Section \ref{subsec_step3_nt}, by searching within the set $\{50, 75, 100, ..., 200\}$ for the value of $\nT$ that minimizes $r^{\rm (mod)}_{\Fxmax}$ with $\mTail=20$. For this task, the score $r^{\rm (mod)}_{\Fxmax}$ was computed with the wave-breaking filter enabled in the stochastic model.

To quantify the effect of the wave-breaking filter in the stochastic model (see Section \ref{sec_enforce_breaking_limit}), the last column of \tab{tab_model_hyperparam} reports the metric:
\begin{equation}
\label{eq_score_extreme_quant}
\dquant =  \frac{\qnnb}{\qnwb} -1  \, ,
\end{equation}
where $\qnnb$ and $\qnwb$ denote the $(1-1/n)$-quantiles predicted by the stochastic model for the response variable $\Fxmax$, with the breaking-filter disabled ("nb") and enabled ("wb"), respectively. 
The effect of the breaking filter is most pronounced for sea states $\#1$ and $\#4$.
Although the DeRisk wave samples for sea states $\#2$, $\#3$, $\#5$, and $\#6$ also contain waves which are near-breaking according to the $\azLag$-criterion, the breaking filter introduced in Section \ref{sec_enforce_breaking_limit} has only a moderate impact on the stochastic model for these sea states. 
This suggests that, for sea states $\#2$, $\#3$, $\#5$, and $\#6$, the stochastic model is able to capture the effect of the DeRisk wave-breaking filter in a purely data-driven manner, whereas this is less the case for sea states $\#1$ and $\#4$.
The DeRisk wave samples corresponding to sea states $\#7$, $\#8$, and $\#9$ do not contain near-breaking waves. Accordingly, no significant effect of the wave-breaking filter was expected, in agreement with the values of $\dquant$ reported in \tab{tab_model_hyperparam}.  

\def\scaleF{0.27}

\def\sizeTTop{0.cm}
\def\sizeTRight{0.0cm}
\def\sizeTBottom{0.0cm}
\def\sizeTLeft{1.8cm}
\def\hshift{-0.4cm}

\def\prefPan{S1L0_NF3100_nH050_mM1_iE132_WB06_simT_}

\def\prefPanT{S7L0_NF1400_nH125_mM1_iE115_WB06_simT_}

\begin{figure}[!h]
\begin{center}
\vspace{-1.5cm}
\begin{tabular}{ccccc} 
    & & Sea State $\#1$ & & \\	
    \begin{overpic}[height=\scaleF\textheight]{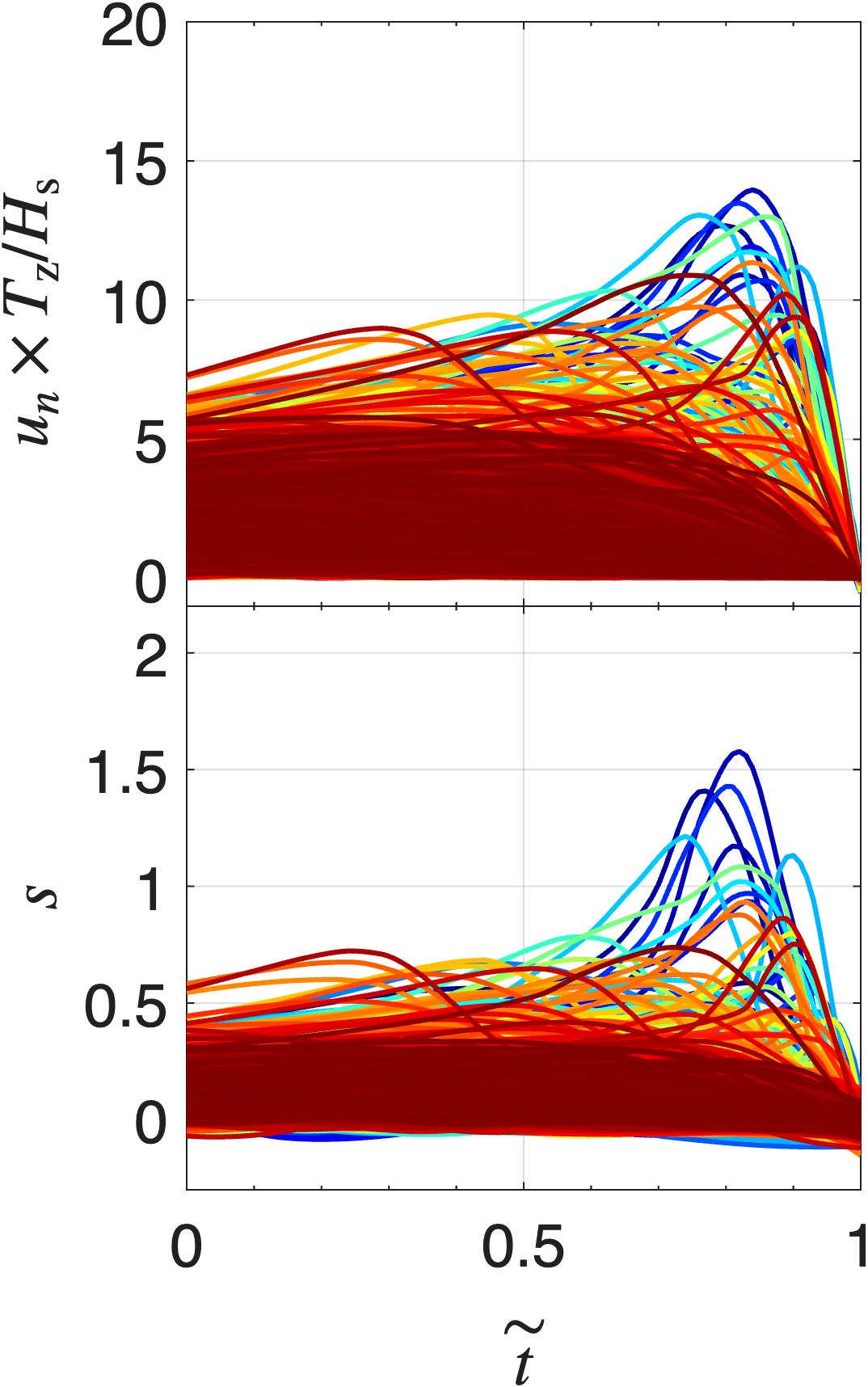}  
    \put(16,92){true sample}
    \end{overpic} &
	\hspace{\hshift} \includegraphics[clip=true, trim={\sizeTLeft{} \sizeTBottom{} \sizeTRight{} \sizeTTop{}}, height=\scaleF\textheight]{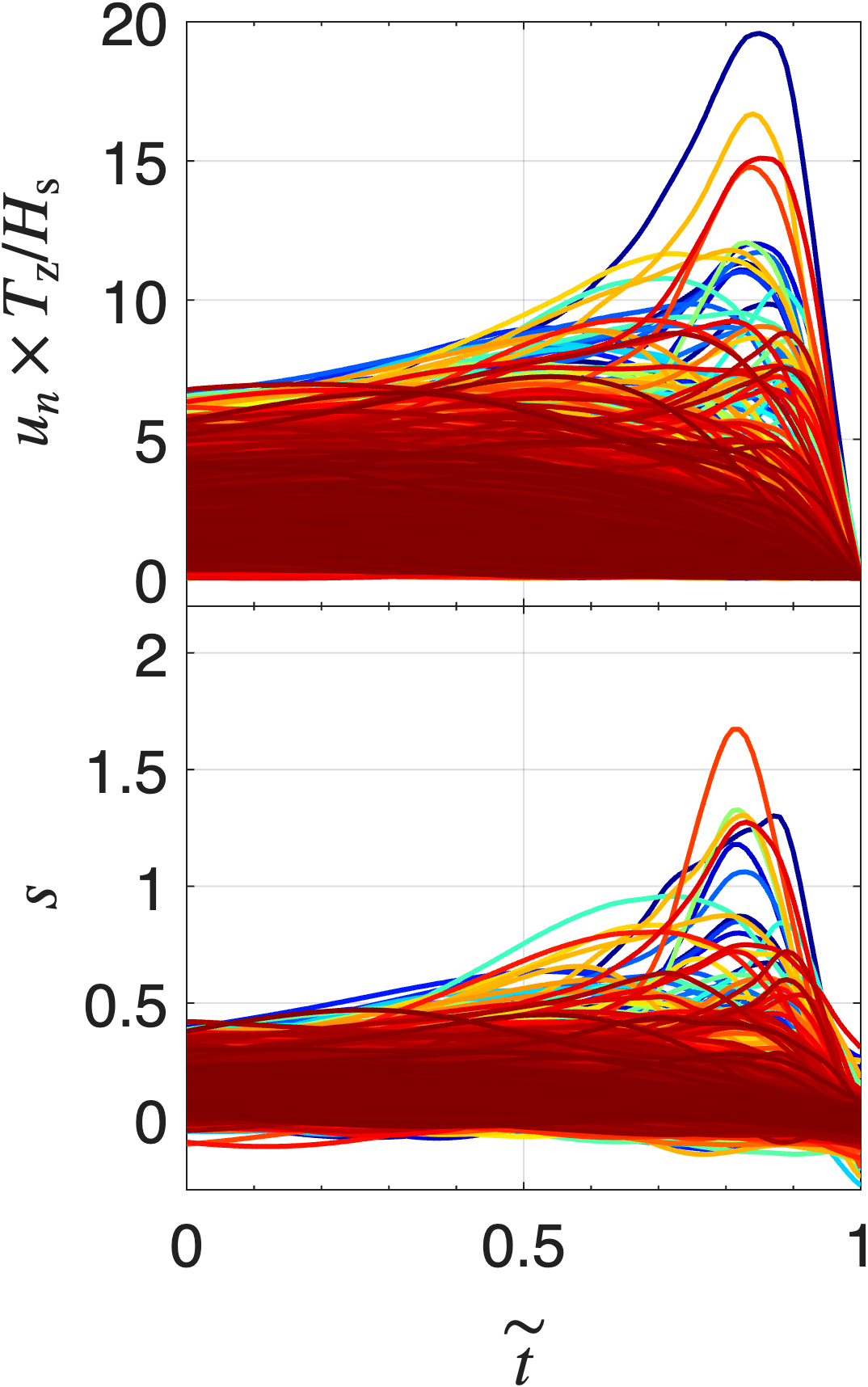} &	
	\hspace{\hshift} \includegraphics[clip=true, trim={\sizeTLeft{} \sizeTBottom{} \sizeTRight{} \sizeTTop{}}, height=\scaleF\textheight]{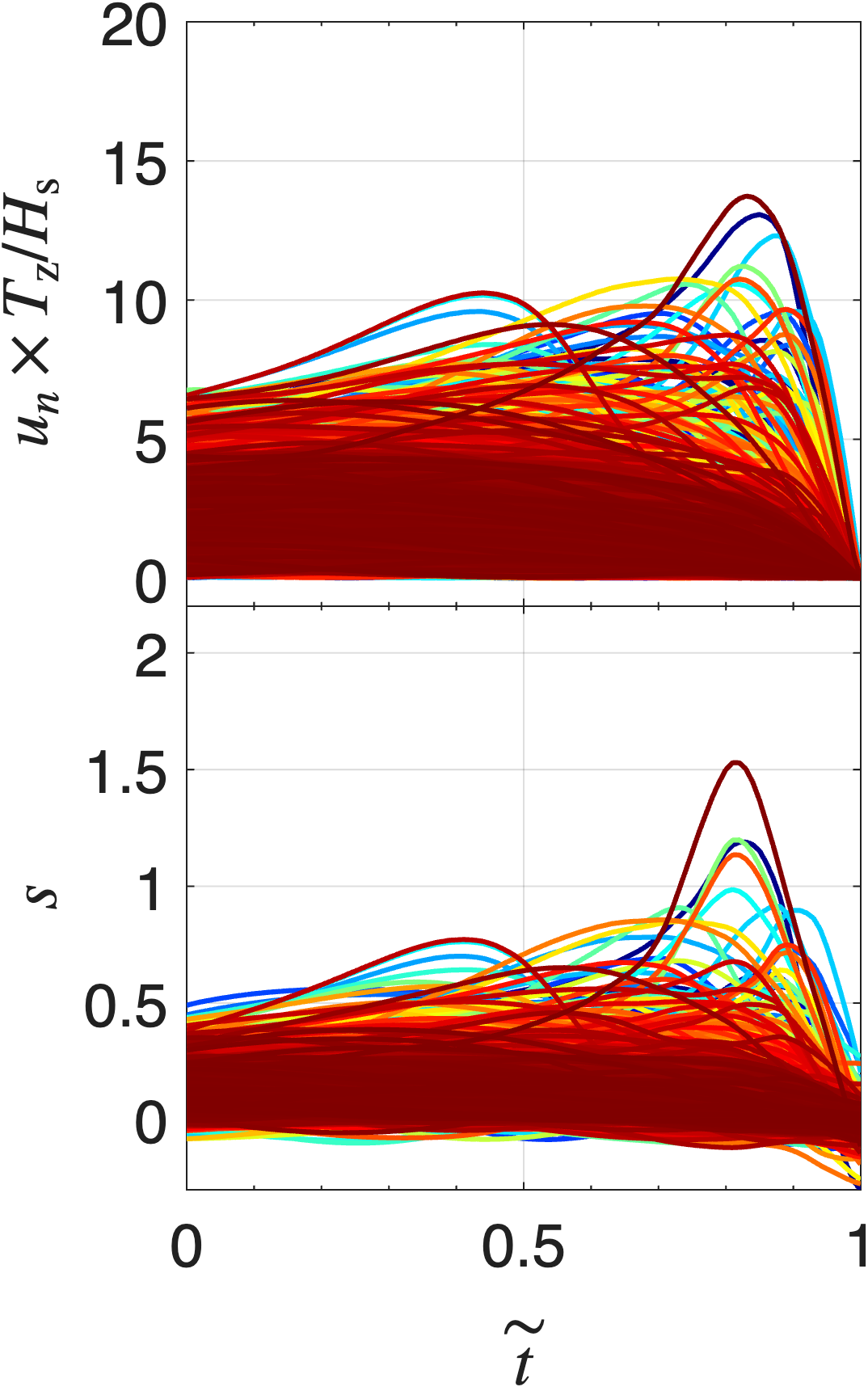} &	
	\hspace{\hshift} \includegraphics[clip=true, trim={\sizeTLeft{} \sizeTBottom{} \sizeTRight{} \sizeTTop{}}, height=\scaleF\textheight]{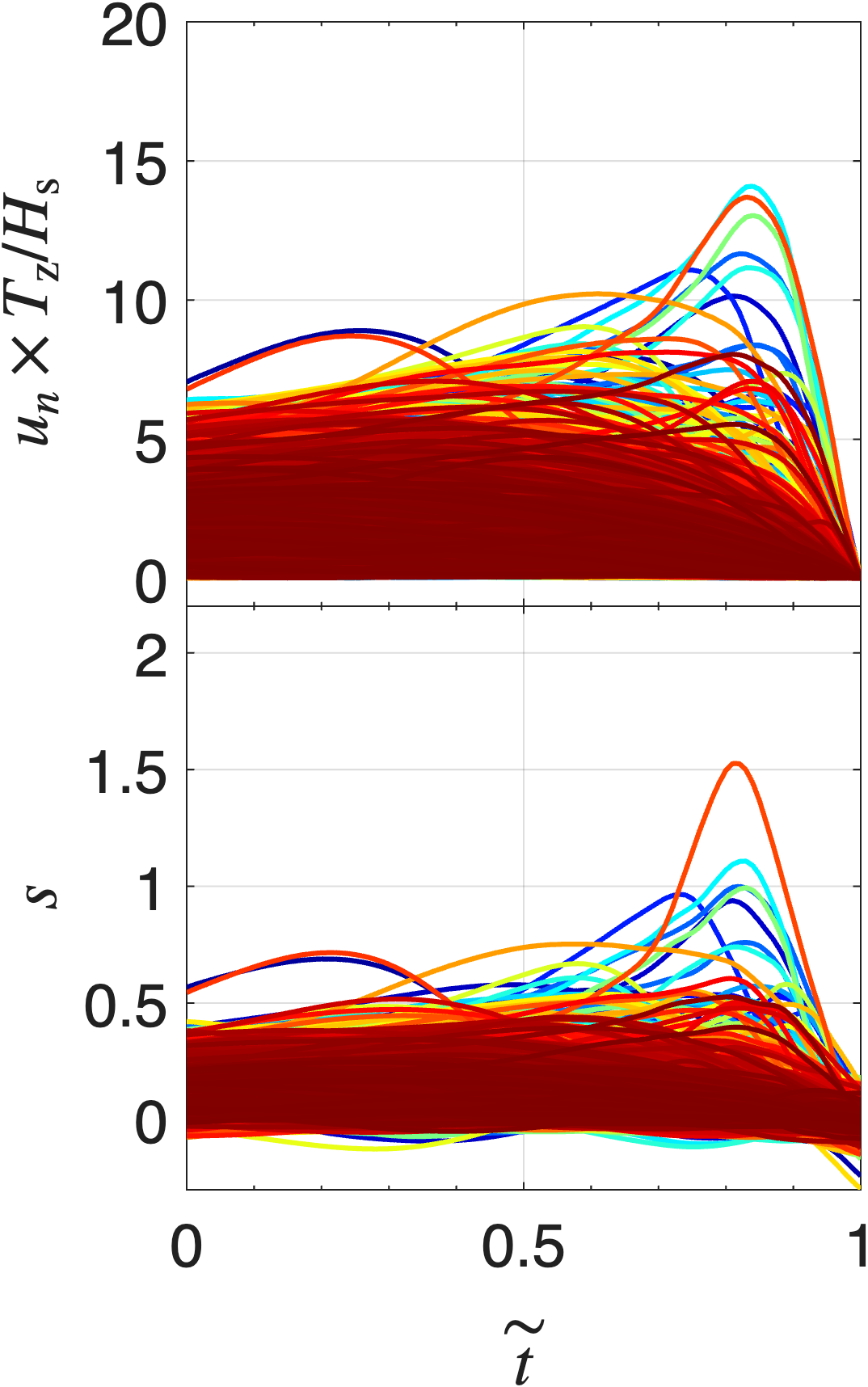} &	
	\hspace{\hshift} \includegraphics[clip=true, trim={\sizeTLeft{} \sizeTBottom{} \sizeTRight{} \sizeTTop{}}, height=\scaleF\textheight]{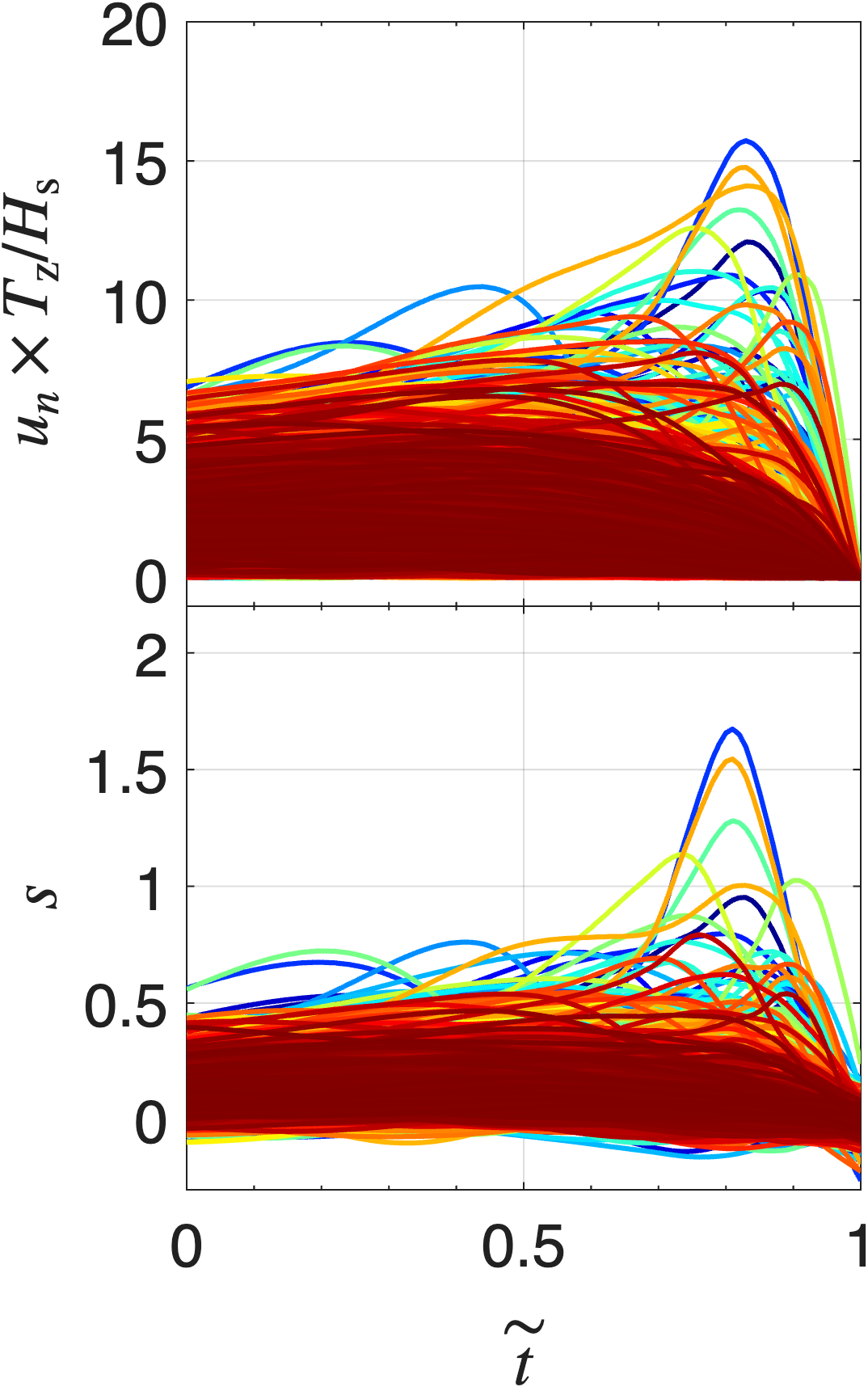}
    \\
    \hline
        & & Sea State $\#7$ & & \\	
    \begin{overpic}[height=\scaleF\textheight]{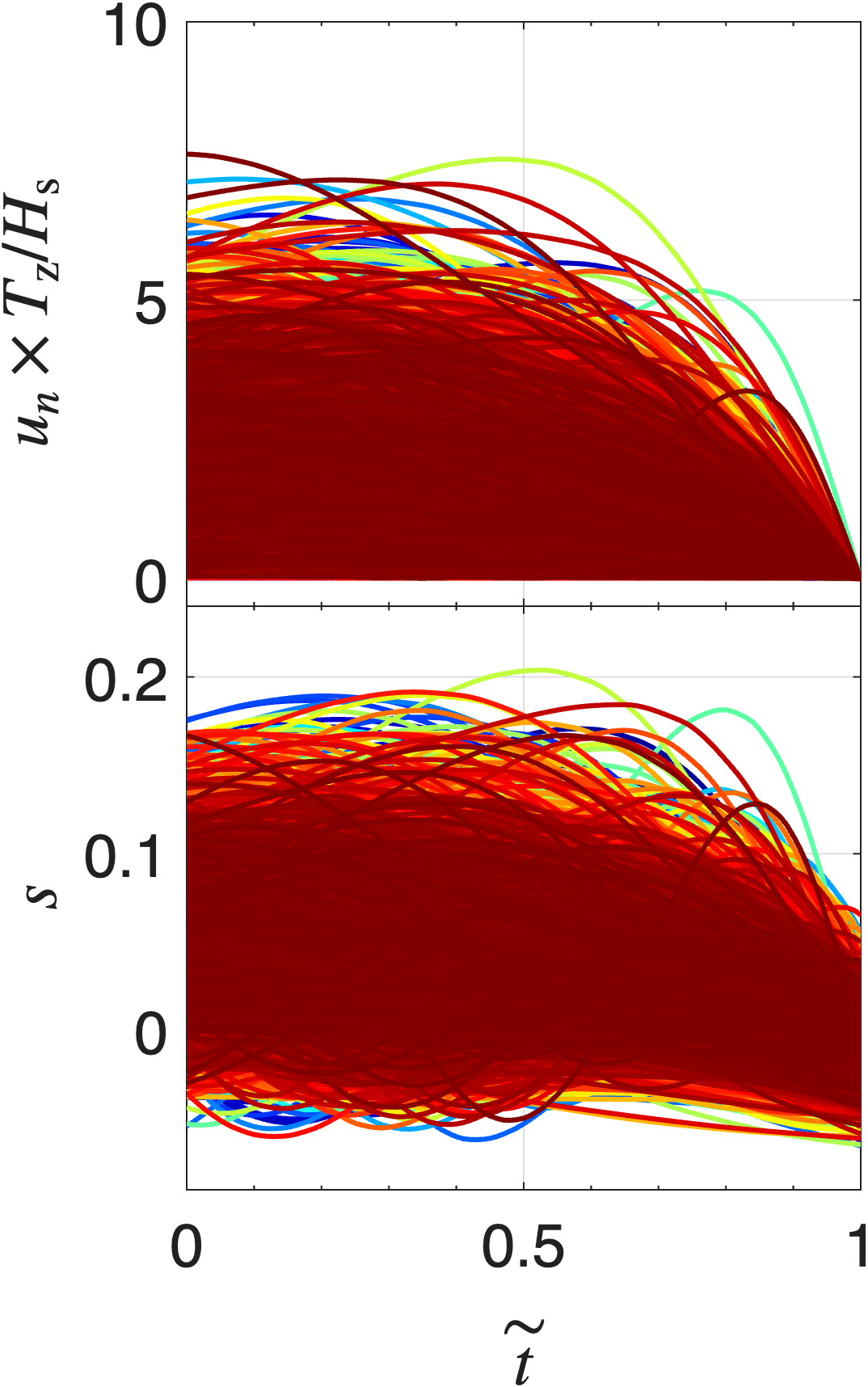}  
    \put(16,92){true sample}
    \end{overpic} &    
	\hspace{\hshift} \includegraphics[clip=true, trim={\sizeTLeft{} \sizeTBottom{} \sizeTRight{} \sizeTTop{}}, height=\scaleF\textheight]{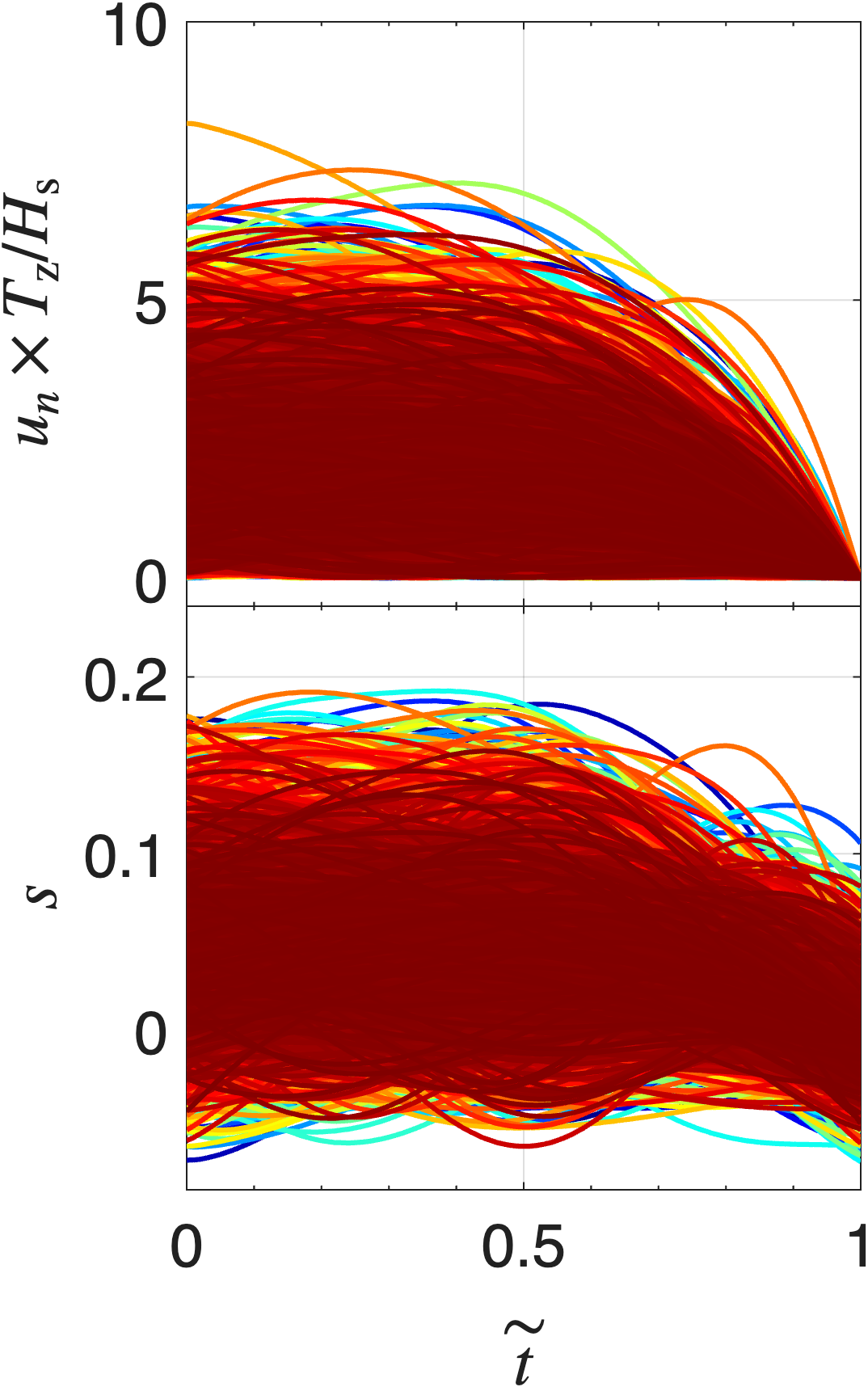} &	
	\hspace{\hshift} \includegraphics[clip=true, trim={\sizeTLeft{} \sizeTBottom{} \sizeTRight{} \sizeTTop{}}, height=\scaleF\textheight]{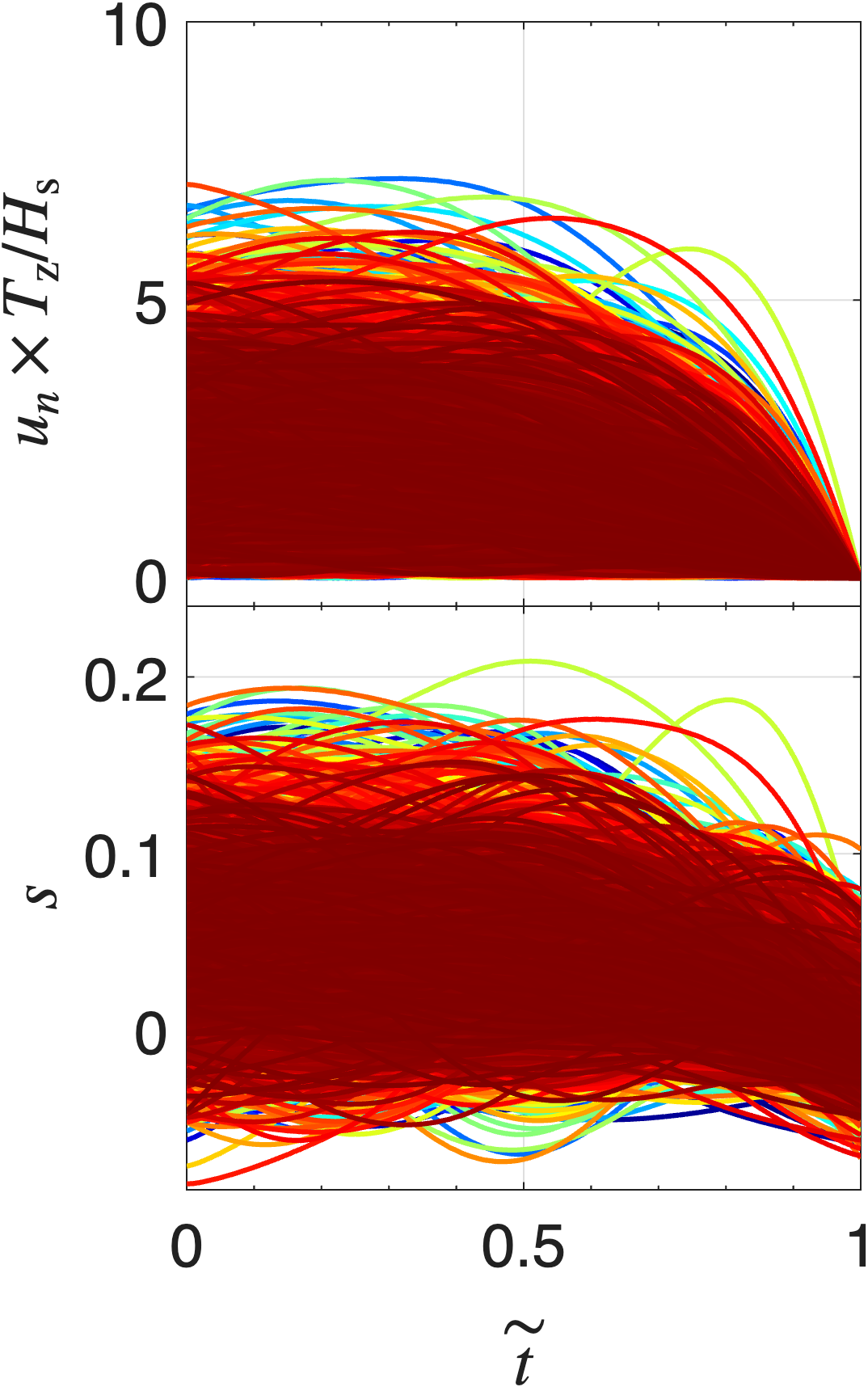} &	
	\hspace{\hshift} \includegraphics[clip=true, trim={\sizeTLeft{} \sizeTBottom{} \sizeTRight{} \sizeTTop{}}, height=\scaleF\textheight]{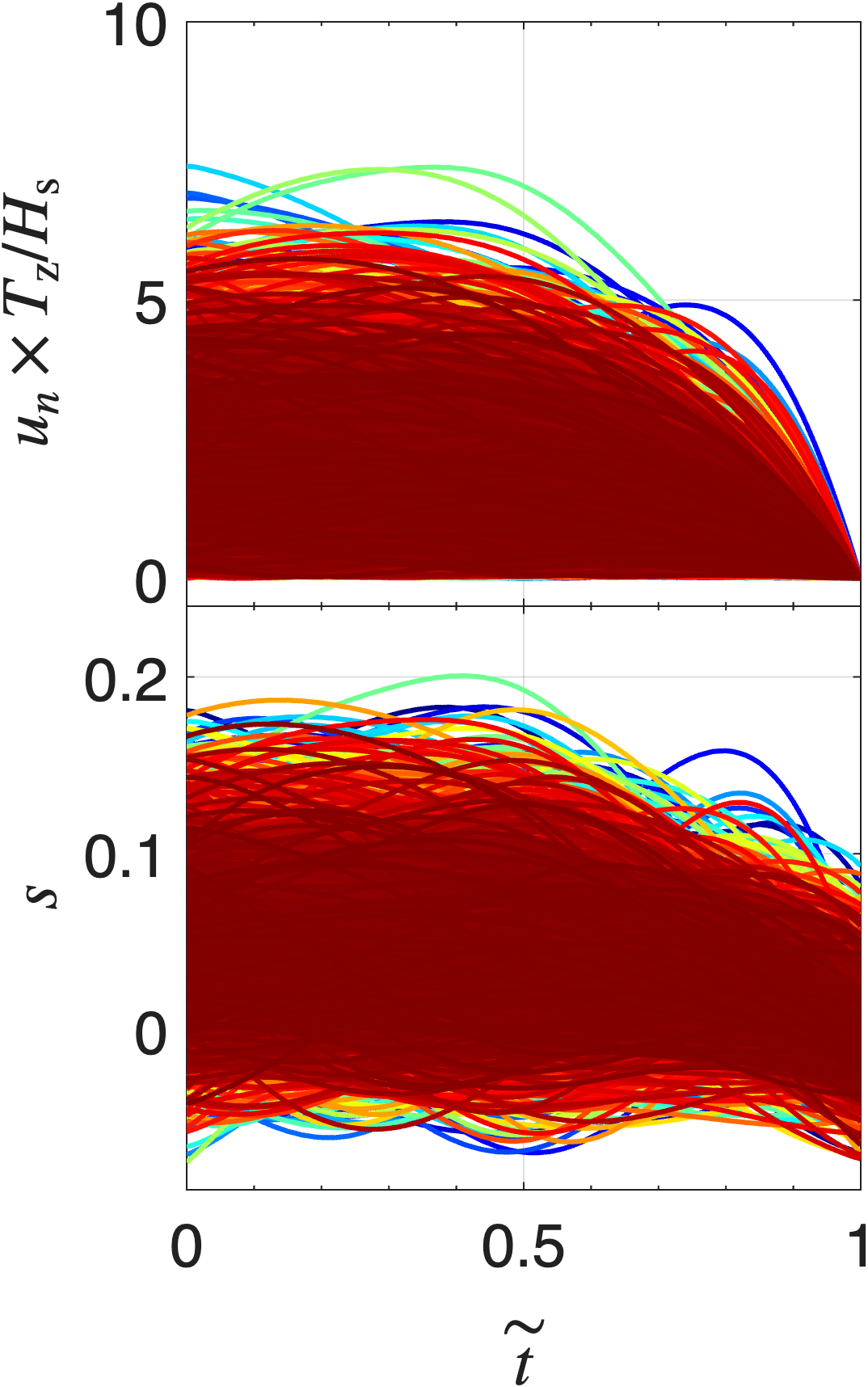} &	
	\hspace{\hshift} \includegraphics[clip=true, trim={\sizeTLeft{} \sizeTBottom{} \sizeTRight{} \sizeTTop{}}, height=\scaleF\textheight]{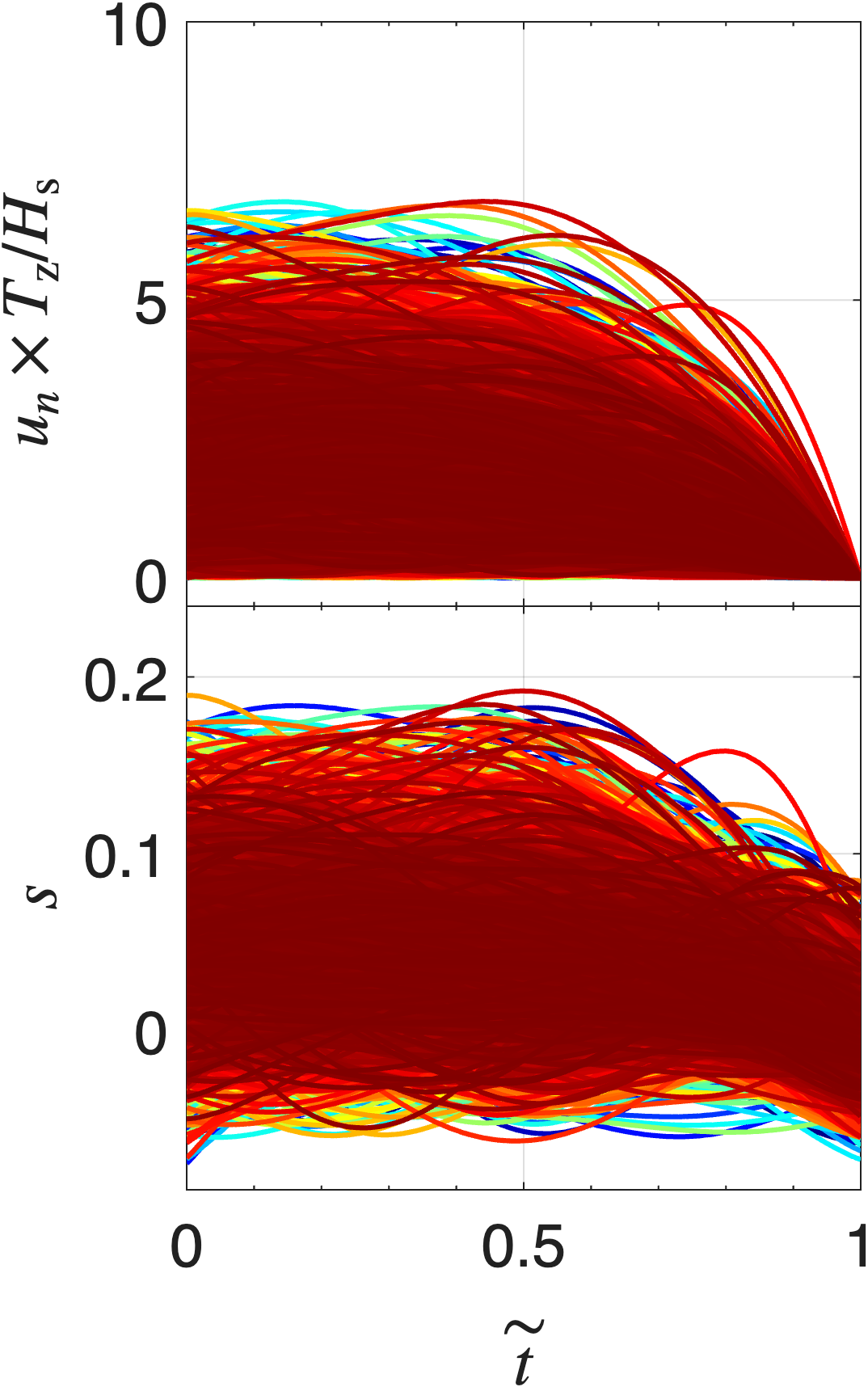}     
\end{tabular}        
\end{center}
\caption{
Samples of stochastic trajectories for $\un$ and $\sx$. The first row shows samples for sea state $\# 1$, while
the second row shows samples for sea state $\#7$.
The trajectory samples extracted from the \derisk{} dataset are labeled as "true sample".
The remaining samples 
are independent synthetic samples generated by the stochastic model.
Each synthetic sample contains the same number of trajectories as the corresponding true sample.
The wave breaking filter is enabled in the stochastic model.
In each plot, colors are used to help distinguish trajectories: 
for a given realization, the trajectories of $\un$ and $\sx$ share the same color.
The corresponding colormap was used from dark blue to dark red, resulting in the bulk of each sample being dominated by dark red.
}
\label{fig_sim_trajs_series}
\end{figure}

\subsection{Stochastic trajectories}
\label{subsec_stoch_trajs}

Fig. \ref{fig_sim_trajs_series} shows  samples of joint stochastic trajectories of $(\un, \sx)$ for sea states $\#1$ and $\#7$. Colors indicate the pairing of $\un$ and $\sx$ trajectories for each realization. 
The trajectory samples extracted from the \derisk{} dataset are labeled as "true sample".
The remaining samples are synthetic samples generated by the model.
Synthetic trajectories were obtained by (i) sampling the feature vector $\Xv$ using Algorithm 1 and (ii) reconstructing trajectories via Eqs. (\ref{eq_pca_reconstruction}-\ref{eq_twe_rec}).\footnote{
When simulating a stochastic trajectory using Algorithm 1 and \eq{eq_pca_reconstruction}, there is a risk that the $\un$-trajectory crosses zero before $\tadim = 1$. This would contradict the fact that the trajectory represents a water-entry event, during which $\un$ remains positive. If such a trajectory is drawn, it is rejected and a new candidate trajectory is simulated.
} 
Each synthetic sample contains the same number of trajectories as the corresponding true sample and was generated with the breaking filter enabled.
The series of synthetic samples in Fig.~\ref{fig_sim_trajs_series} illustrates the variability of trajectories produced by the model.

The stochastic model is able to generate synthetic trajectories that resemble the true trajectories. In particular, for sea state $\#1$, the model captures the tendency of extreme trajectories to exhibit a peak around $\tadim \simeq 0.7-0.9$.
These trajectories correspond to waves with steep slopes near the crest, which also happen to be the highest waves in the simulated sea state (see Fig.~\ref{fig_wave_extraction}, top right panel). 
In contrast, for sea state $\#7$, extreme trajectories do not appear to differ significantly in shape from the rest of the population. 

The fact that extreme trajectories have shapes distinctly different from the bulk of the population in sea state $\#1$, while this is not the case in sea state $\#7$, explains why the number of principal components retained is higher for sea state $\#1$ than for sea state $\#7$ (see \tab{tab_model_hyperparam}).

\def\scaleF{0.38}
\def\sizeF{3.9cm}
\def\sizeT{1.23cm}
\def\sizeS{-0.4cm}
\def\sizeSS{-0.4cm}

\begin{figure}[h!]
\begin{center}
\begin{tabular}{ccccc} 
	\hspace{\sizeSS{}}\centering \begin{turn}{90} \hspace{0.8cm} {\Large sea state \#1}\end{turn}  &
	\includegraphics[clip=true, trim={0 0 0 0}, height=\sizeF]{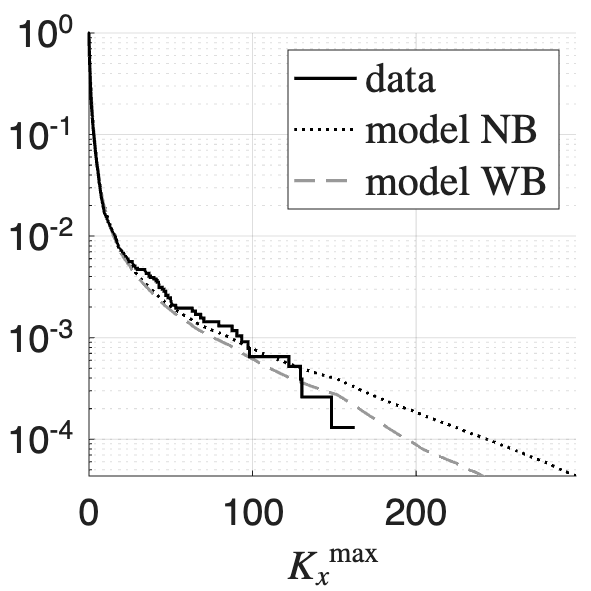} &
	\hspace{\sizeS{}}\includegraphics[clip=true, trim={\sizeT{} 0 0 0}, height=\sizeF]{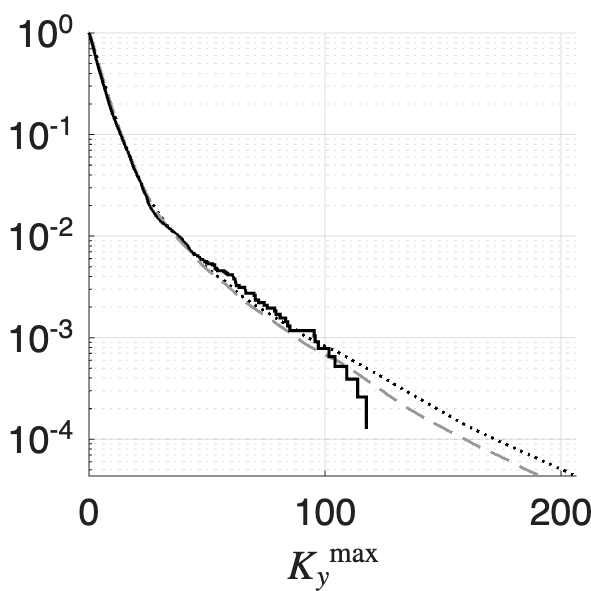} &
	\hspace{\sizeS{}}\includegraphics[clip=true, trim={\sizeT{} 0 0 0}, height=\sizeF]{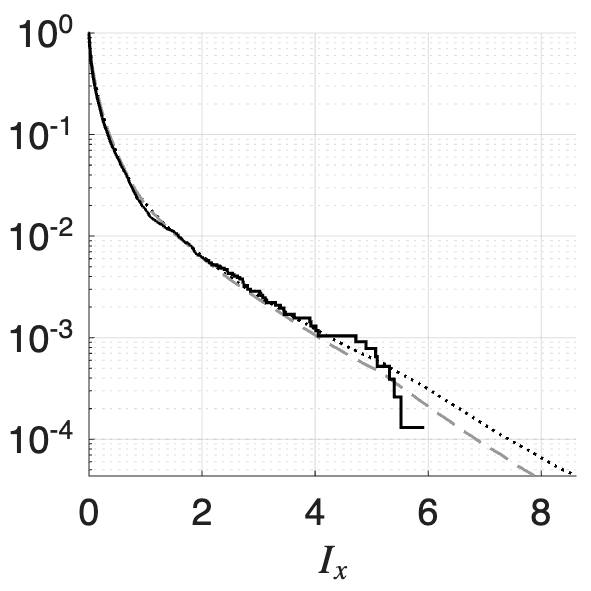} &	
	\hspace{\sizeS{}}\includegraphics[clip=true, trim={\sizeT{} 0 0 0}, height=\sizeF]{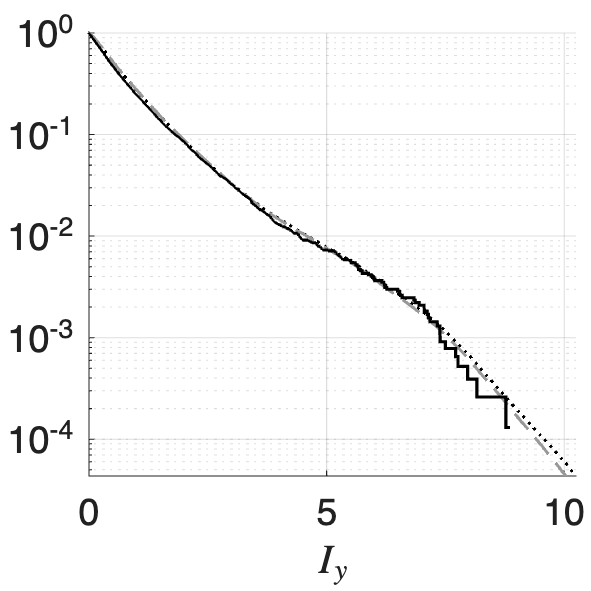} \\
	\hspace{\sizeSS{}}\centering \begin{turn}{90} \hspace{0.8cm} {\Large sea state \#7}\end{turn}  &
	\includegraphics[clip=true, trim={0 0 0 0}, height=\sizeF]{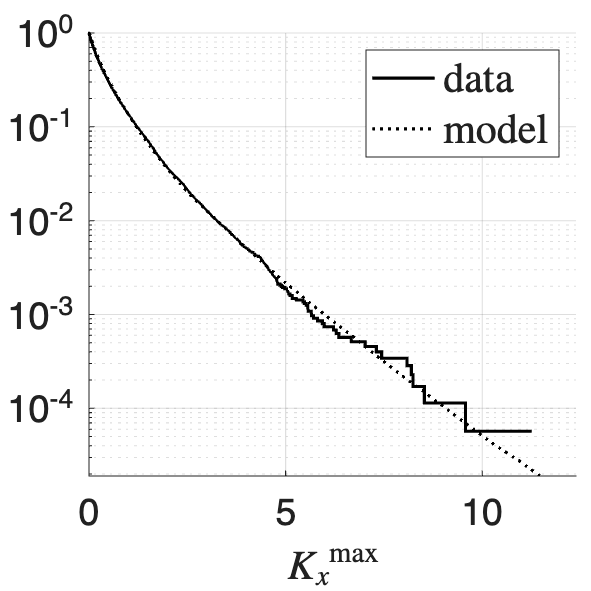} &
	\hspace{\sizeS{}}\includegraphics[clip=true, trim={\sizeT{} 0 0 0}, height=\sizeF]{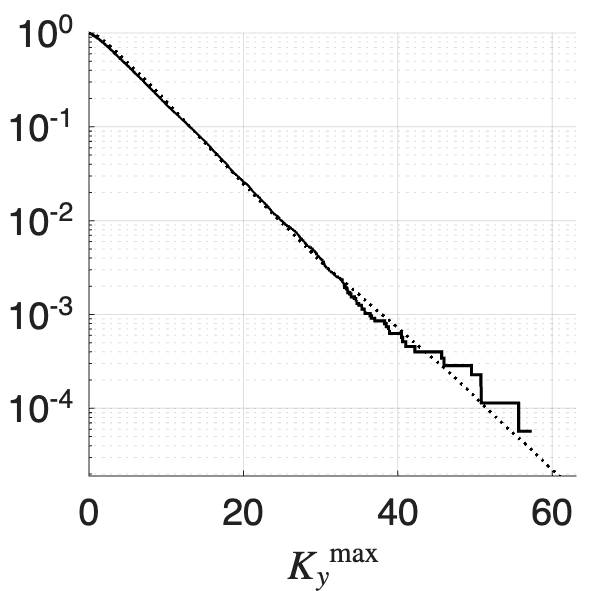} &
	\hspace{\sizeS{}}\includegraphics[clip=true, trim={\sizeT{} 0 0 0}, height=\sizeF]{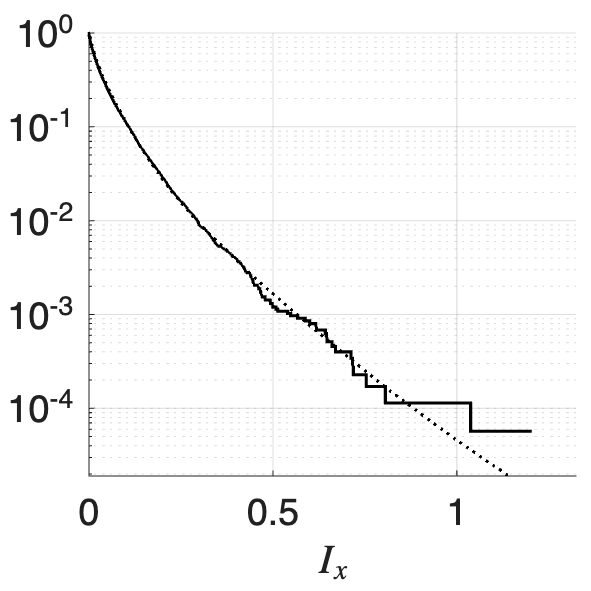} &	
	\hspace{\sizeS{}}\includegraphics[clip=true, trim={\sizeT{} 0 0 0}, height=\sizeF]{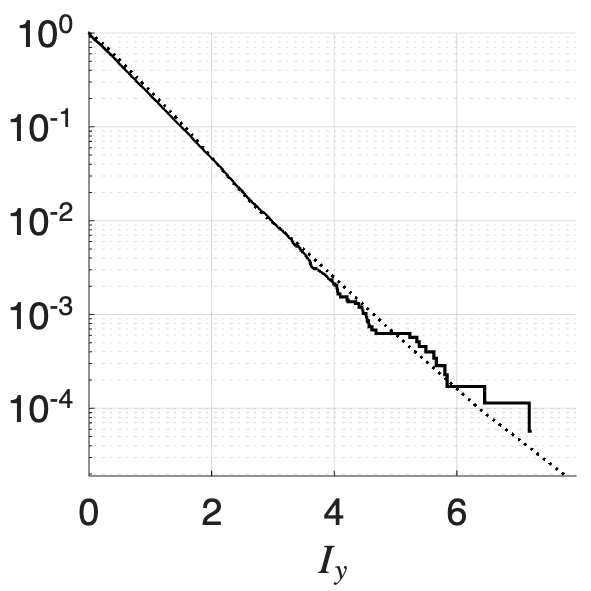} 	
\end{tabular}        
\end{center}
\caption{
Marginal distributions of the response variables, $\Fxmax$, $\Fymax$, $\Ix$, and $\Iy$, shown as exceedance probabilities. 
The first (resp. second) row shows results for sea state \#1 (resp. \#7).
For sea state \#1, two versions of the stochastic model are shown: one version with the breaking filter enabled (model WB), and one version with the breaking filter disabled (model NB). 
For sea state \#7, enabling the breaking filter has effectively no effect; 
consequently, only one curve is shown for the model.
}
\label{fig_dists_response}
\end{figure}

\subsection{Distributions of the response variables}
\label{subsec_dist_resp_vars}

\fig{fig_dists_response} shows the marginal distributions of the response variables defined 
in Eqs. 
(\ref{eq_kxmax}--\ref{eq_iy}),
for sea states $\#1$ and $\#7$.
The empirical distributions (labeled as "data" in the legend) were obtained from the "true samples" of trajectories shown in Fig.~\ref{fig_sim_trajs_series}.
Model distributions were obtained by generating numerous synthetic trajectories of 
$\un$ and 
$\sx$ from the stochastic model and computing the response variables via Eqs. 
(\ref{eq_kxmax}--\ref{eq_iy}).
An importance sampling scheme was employed to reduce the number of realizations required to achieve satisfactory precision in the tails of the distributions.
For sea state $\#1$, two versions of the stochastic model are shown: one with the wave-breaking filter enabled (labeled "WB" in the legend), following the approach described in Section \ref{sec_enforce_breaking_limit}, and one without the filter (labeled "NB").
The effect of the breaking filter is most pronounced for the response variable $\Fxmax$.
For sea state $\#7$, the wave-breaking filter has a negligible effect; consequently, only one version of the model is reported.

The model distributions show good agreement with the empirical distributions obtained from the \derisk{} dataset for both sea states $\#1$ and $\#7$: 
the observed differences remain within the range of expected sampling variability.
The distribution tails of 
$\Fxmax$ and $\Fymax$
 are much heavier in sea state $\#1$ than in sea state $\#7$. 
 This reflects the fact that, in sea state $\#1$, extreme waves tend to have shapes that are distinct from the bulk of the population.
Moreover, in sea state $\#1$,
the tails of the distributions of $\Ix$ and $\Iy$ are lighter than those of  
$\Fxmax$ and $\Fymax$, respectively. 
This follows from the fact that trajectories producing extreme values of $(\un, \sx)$ correspond to the steepest waves, which are also characterized by relatively short rise times to their crests (see \fig{fig_wave_extraction}).
These shorter durations counterbalance the high peak values of $\Fx$ and $\Fy$ when computing $\Ix$ and $\Iy$ via Eqs. (\ref{eq_ix}-\ref{eq_iy}).

\def\scaleF{0.39}
\def\sizeS{-0.3cm}
\def\sizeSS{-0.4cm}

\def\prefBoot{bootstrap_S1L0_NF3100_nH050_mM1_iE132_WB06_}

\def\prefBootBis{bootstrap_S7L0_NF1400_nH125_mM1_iE115_WB06_}

\begin{figure}[t!]
\begin{center}
\begin{tabular}{ccccc} 
	\hspace{\sizeSS{}}\centering \begin{turn}{90} \hspace{0.5cm} {\large sea state \#1}\end{turn}  &
	\includegraphics[scale=\scaleF]{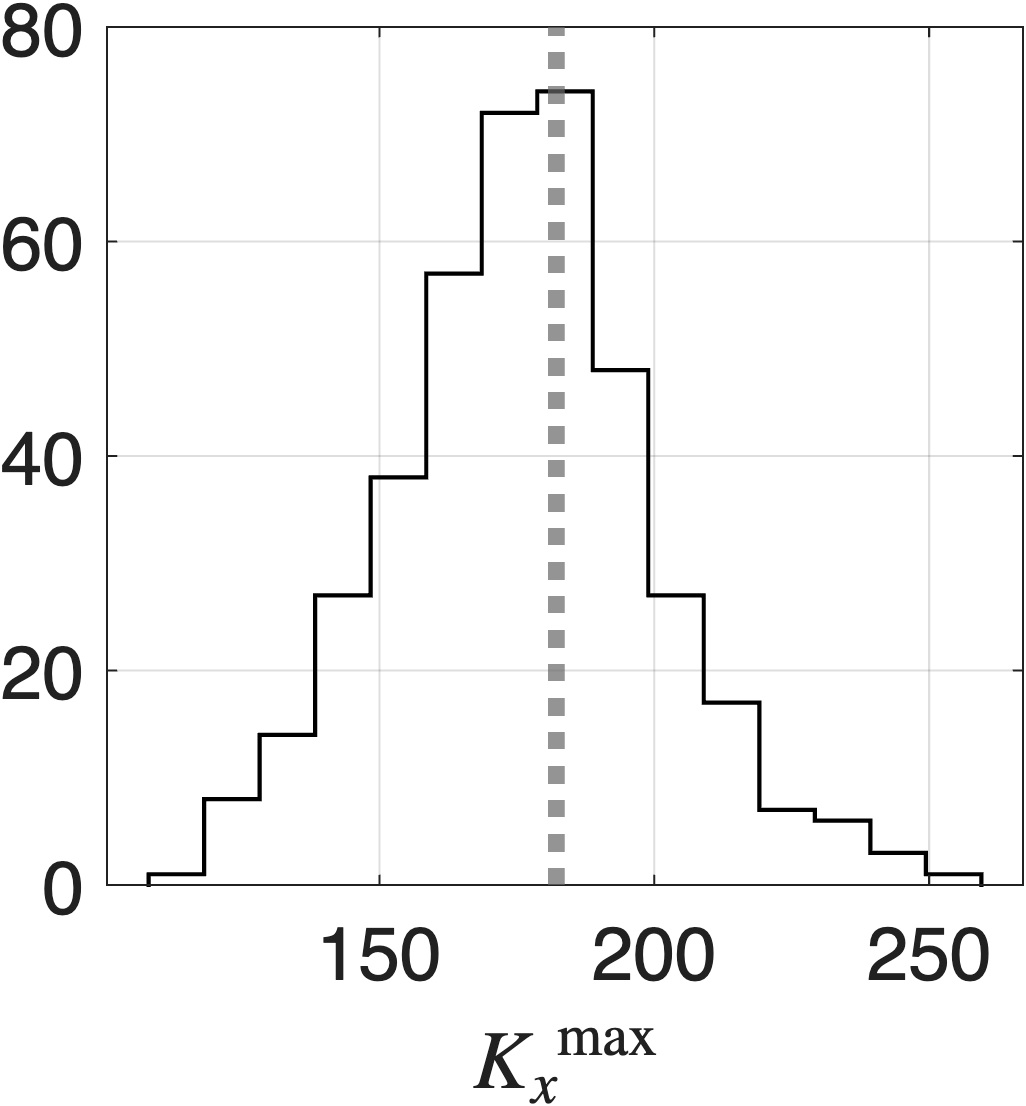} &
	\hspace{\sizeS{}}\includegraphics[scale=\scaleF]{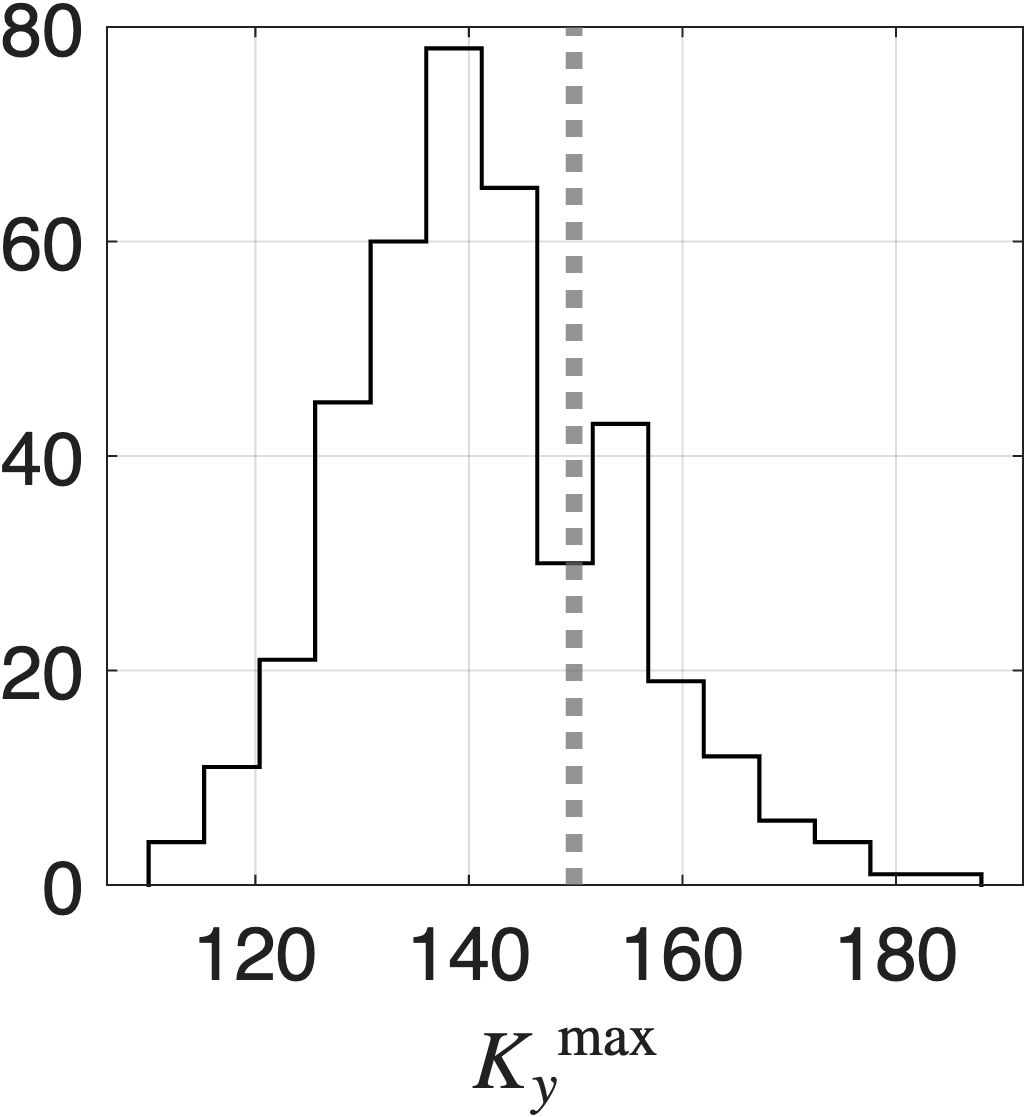} &
	\hspace{\sizeS{}}\includegraphics[scale=\scaleF]{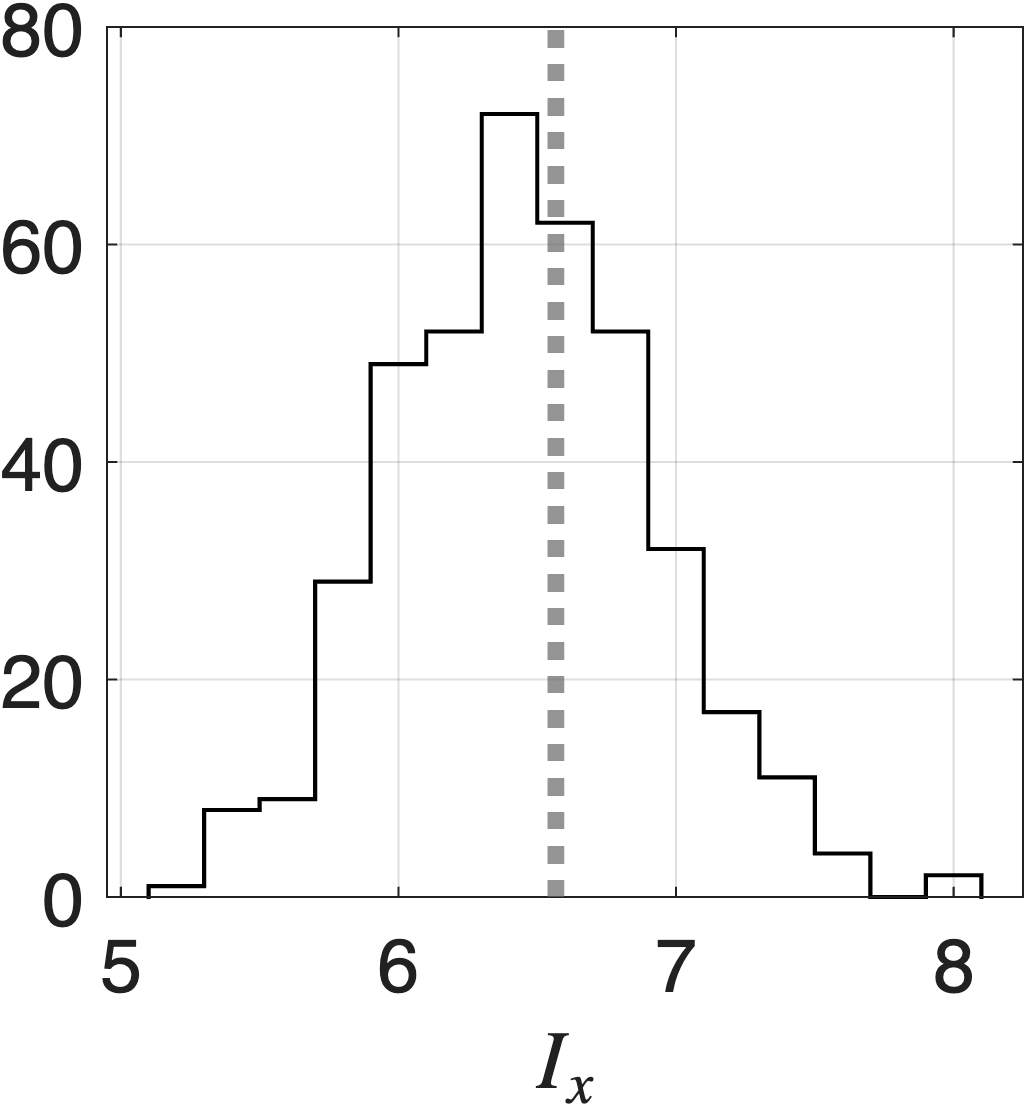} &	
	\hspace{\sizeS{}}\includegraphics[scale=\scaleF]{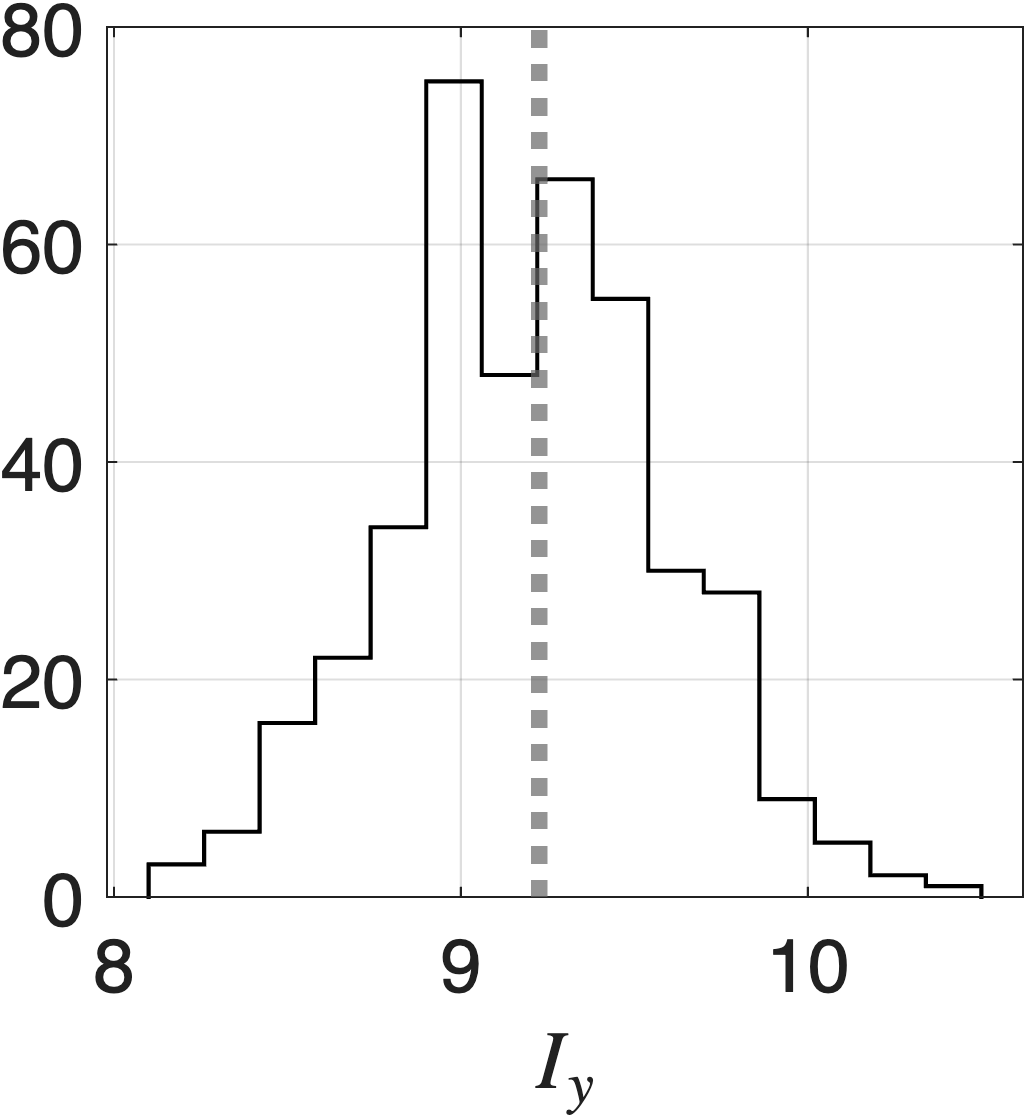} \\
	\hspace{\sizeSS{}}\centering \begin{turn}{90} \hspace{0.5cm} {\large sea state \#7}\end{turn}  &
	\includegraphics[scale=\scaleF]{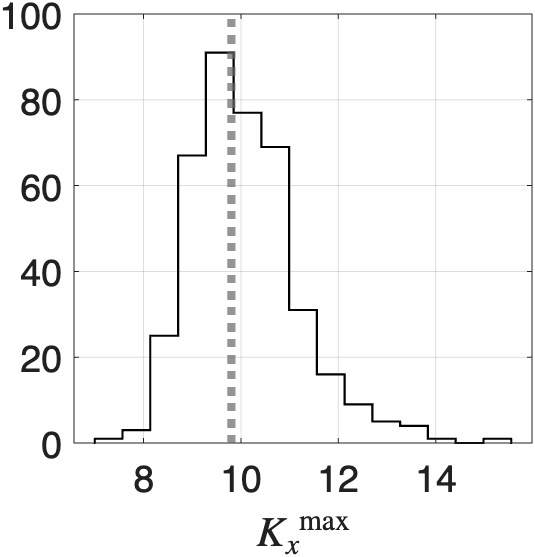} &
	\hspace{\sizeS{}}\includegraphics[scale=\scaleF]{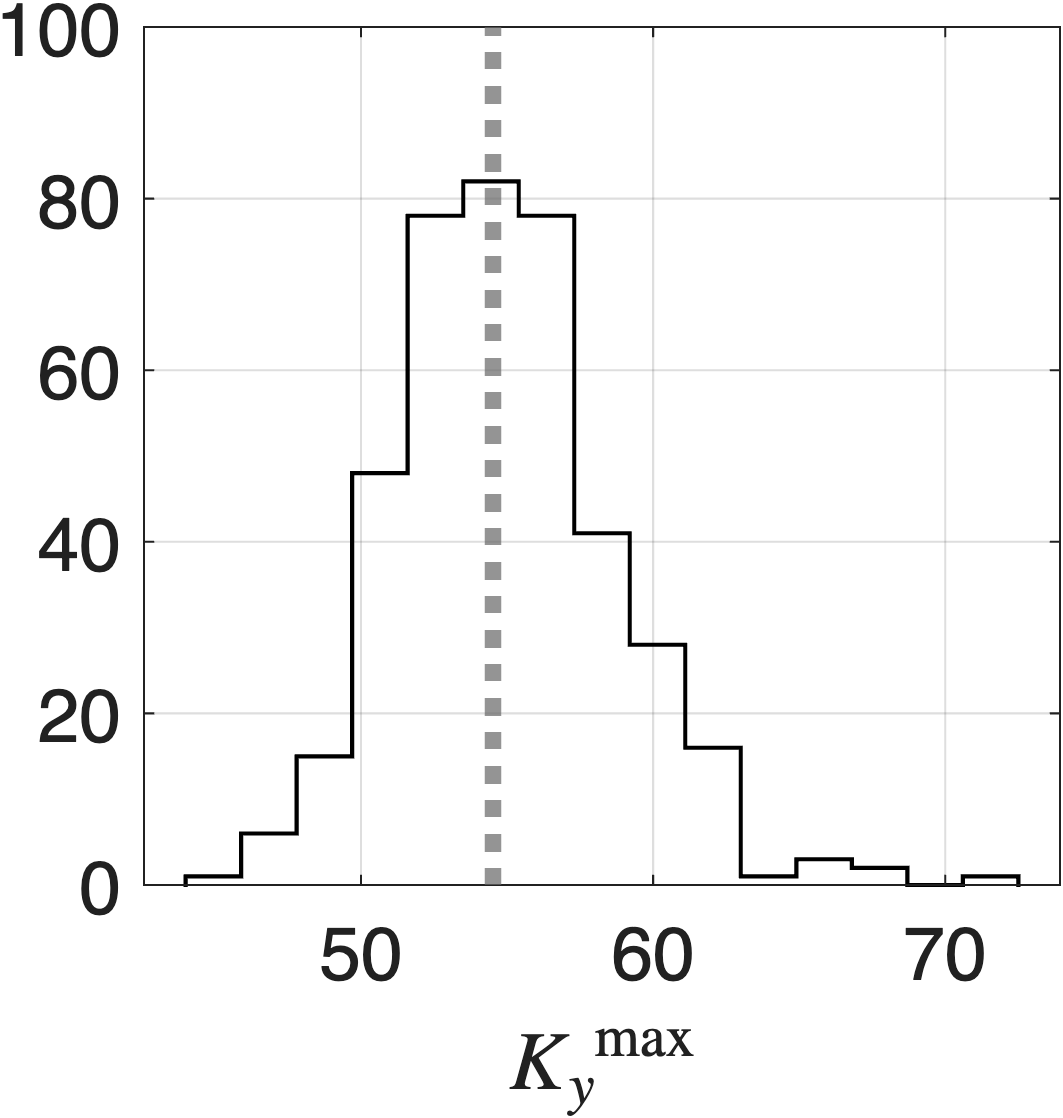} &
	\hspace{\sizeS{}}\includegraphics[scale=\scaleF]{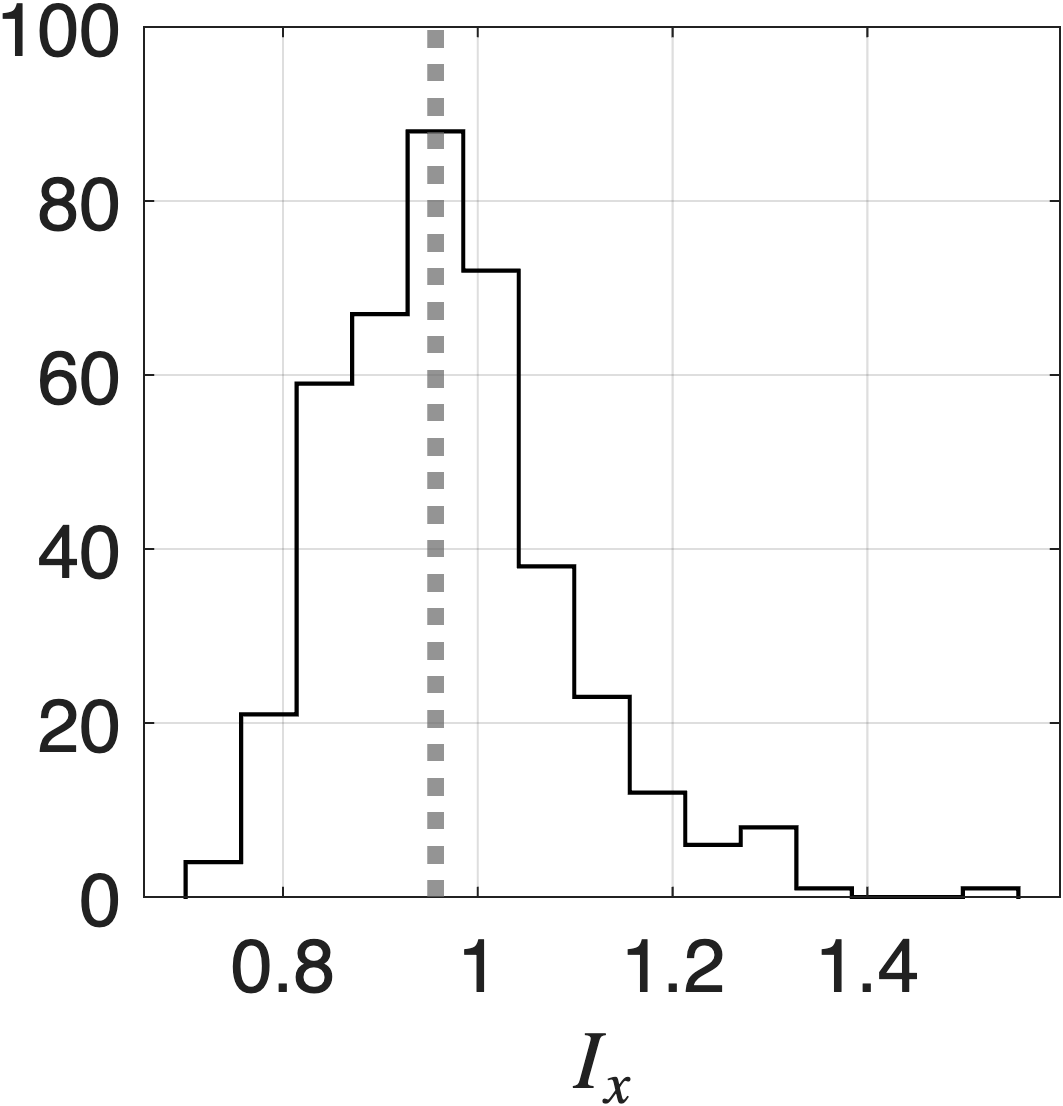} &	
	\hspace{\sizeS{}}\includegraphics[scale=\scaleF]{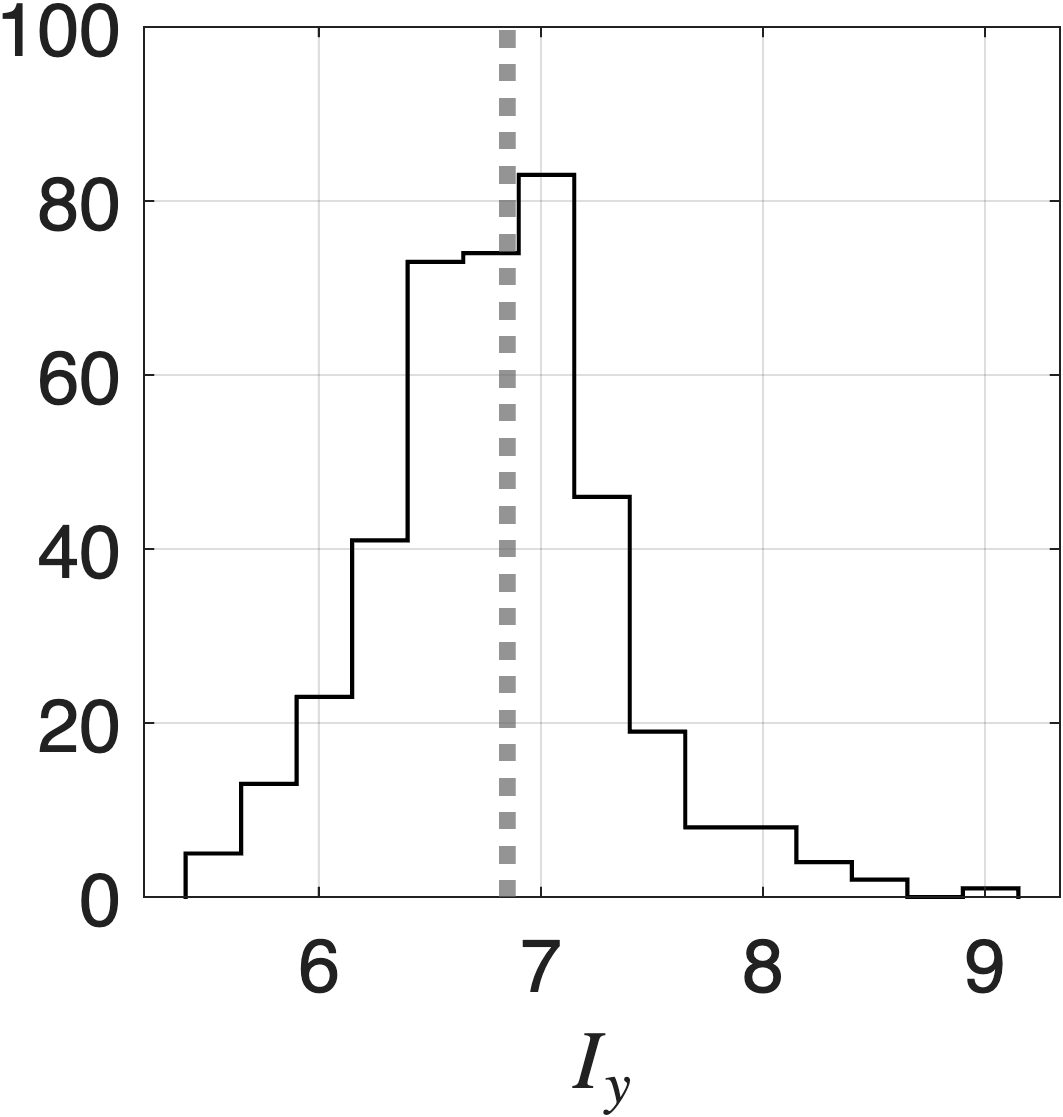} 	
\end{tabular}        
\end{center}
\caption{
Bootstrap histograms of the $(1-1/\nobs)$-quantiles of the response variables. 
The first and second rows show results for sea state $\#1$  
and sea state $\#7$, respectively. 
The wave-breaking filter was enabled.
The histograms were generated from $400$ bootstrap samples.
}
\label{fig_bootstrap}
\end{figure}

\subsection{Model sensitivity to dataset variations}
\label{subsec_model_sensitivity}

To test the sensitivity of the model to fluctuations in the dataset and to assess the variance of its predictions, a bootstrap approach was implemented.
Results are shown only for sea states $\#1$ and $\#7$; comparable levels of sensitivity were observed for other sea states.
The analysis focuses on the tails of the predicted distributions, as statistics associated with extreme events are typically more sensitive to dataset fluctuations than statistics characterizing the bulk of the distribution.
In the context of extreme value analysis, several approaches have been proposed to construct confidence intervals using bootstrap techniques \cite{caers_1998, gomes_2001, wang_2010, gilleland_2020}. Here, we adopt a nonparametric bootstrap approach in which the dataset is resampled with replacement to generate bootstrap samples of the same size $\nobs$ as the original dataset.
Each bootstrap dataset is then fitted with the stochastic model using the same hyperparameters as in the original setting (see \tab{tab_model_hyperparam}).

\fig{fig_bootstrap} shows the bootstrap histograms obtained for the $(1 - 1/\nobs)$-quantiles of the response variables. 
The wave-breaking filter was enabled.
The values computed from the original dataset are indicated by vertical dotted lines.
Across the bootstrap distributions shown in \fig{fig_bootstrap}, the ratio of the standard deviation to the mean ranges from approximately $0.05$ to $0.13$.

\section{Summary \& Discussion}
\label{sec_discussion}

This study has formulated and investigated a data-driven approach for the stochastic modeling of water waves in a given sea state, with the \derisk{} database serving as a case study.
The illustrative examples focus on modeling water-entry stochastic trajectories. 
The kinematic variables considered are the slope of the free surface ($\sx$), the component of the fluid velocity normal to the free surface ($\un$), and the vertical Lagrangian acceleration of the fluid at the free surface ($\azLag$). The latter has been included to enforce a wave-breaking limit within the stochastic model.
Nine representative sea states covering the \derisk{} parameter space were investigated. Detailed results on the stochastic trajectories and the distributions of selected response variables were presented for a strongly nonlinear sea state ($\#1$) and a weakly nonlinear sea state ($\#7$).
As illustrated, the model can effectively simulate waves for a given sea state, reproducing both the bulk of the population and the extreme wave population.
In particular, the model is able to capture the smooth transition from the bulk of the population to the class of extreme waves.
This makes the approach potentially valuable for reliability analysis, for example, when investigating the probability distribution of loads induced by waves on a marine structure.

The use of a simplified vine copula for the bulk of the distribution, combined with the Heffernan and Tawn (HT) approach for the multivariate tail, makes the model readily scalable in terms of dimensionality. 
An alternative approach was proposed by Naveau et al. (2016) \cite{naveau_2016}, which models the distribution “continuously” without explicitly distinguishing between the bulk and the tail. 
This method is attractive due to its parsimony in parameters and the elimination of threshold selection.
However, it has been demonstrated only for univariate \cite{naveau_2016} and bivariate distributions \cite{legrand_2023}, and it is not readily extendable to higher dimensions.

In the illustrative examples, the stochastic model has a dimension ranging from $14$ to $33$, depending on the considered sea state (see \tab{tab_model_hyperparam}).
Despite the relatively high dimensionality of the model, the number of hyperparameters was reduced to three: 
the number of components in the PCA analysis ($\npca$), 
the set of considered HT conditional models ($\seqHT$), and the number of observations used for tail modeling ($\nT$).
A sequential strategy for adjusting these three hyperparameters was proposed 
in Section \ref{seq_strategy_hyperparams}.
Two metrics based on the extreme quantiles of the response variables were used to select $\npca$ and $\nT$.

The proposed statistical model is interpretable to some extent, as it is based on a principal component decomposition of the stochastic trajectories.
In particular, the principal component(s) driving a trajectory to be extreme can be identified. 
In the present application, the score of the first principal component was found to be extreme for most of the extreme trajectories. 
This observation enabled a parsimonious modeling of the multivariate tail by considering only two conditional HT models.

The approach is generic and can be applied to model other multivariate processes. 
As the number of variables increases, additional composite principal components may be needed to faithfully capture the trajectories, which in turn would increase the model’s dimensionality.
However, this increase would not be necessarily substantial, since joint FPCA can exploit correlations between the trajectories of different variables.
For instance, in the examples considered in this paper, the trajectories of $\un$ and $\sx$ exhibit a strong positive correlation.
The method can also be applied to model waves over their entire period, rather than restricting the analysis to the water-entry phase. Extending the analysis to the full wave period may likewise require additional principal components. 
Investigating the practicality of the proposed approach for configurations necessitating higher dimensionality may be the subject of future research.

\section*{Acknowledgments}

The author gratefully acknowledges V. Monbet, P. Naveau, and N. Raillard for insightful discussions on statistical analysis, and F. Pierella for valuable information regarding the DeRisk database.


\setcounter{figure}{0}
\appendix


\section{Simplified vine copula approach}
\label{appendix_vine_copula}

A copula is a multivariate distribution whose marginal distributions are uniform on $[0,1]$.
In the present work, the marginal distributions of the feature vector $\Xv$ are assumed to be continuous.
The copula of $\Xv$ is then defined as the joint cumulative distribution function of the random vector $\Uv = (U_1, U_2, ..., U_d) = (F_1(X_1), F_2(X_2), ..., F_d(X_d))$:
\begin{equation}
C(u_1, u_2, ..., u_d) = \Pr[U_1 \le u_1, U_2 \le u_2, ..., U_d \le u_d] \, .
\end{equation}
Following the vine copula approach \cite{bedford_2001, bedford_2002, aas_2009, kurowicka_2010, nagler_2014}, 
the copula density $c$ can be decomposed as the product of $d(d-1)/2$ bivariate conditional copula densities.
For instance, a $4$-dimensional copula density can be expressed as:
\begin{align}
c_{1234} & (u_1,u_2,u_3,u_4)  =   \nonumber \\
& c_{1,2}(u_1,u_2) \times c_{2,3}(u_2,u_3) \times c_{2,4}(u_2,u_4) \nonumber \\
& \times c_{1,4;2}(C_{1|2}(u_1;u_2),C_{4|2}(u_4;u_2);u_2) \times c_{1,3;2}(C_{1|2}(u_1;u_2),C_{3|2}(u_3;u_2);u_2) \nonumber \\
& \times c_{3,4;1,2}(C_{3|1,2}(u_3;u_1u_2),C_{4|1,2}(u_4; u_1, u_2);u_1, u_2) \label{eq_cop_decomp_nonSimp}  \, ,                               
\end{align}
where:
\begin{itemize}
\item $c_{i,j}$ is the copula density of the pair $(U_i,U_j)$;
\item $C_{i|j}(.;u_j)$ is the conditional distribution function of $U_i$, given $U_j = u_j$;
\item $C_{i|j,k}(.;u_j,u_k)$ is the conditional distribution function of $U_i$, given $U_j = u_j$ and $U_k=u_k$;
\item $c_{i,j;k} (.,.;u_k)$ is the bivariate density of $(U_{i|k},U_{j|k})|U_k = u_k$, with
$U_{i|k} = C_{i|k}(U_i;U_k)$ and $U_{j|k} = C_{j|k}(U_j;U_k)$; 
\item $c_{i,j;k,l}(.,.;u_k,u_l)$ is the bivariate density of $(U_{i|k,l},U_{j|k,l})|(U_k = u_k,U_l = u_l)$, with
$U_{i|k,l} = C_{i|k,l}(U_i;U_k,U_l)$ and $U_{j|k,l} = C_{j|k,l}(U_j;U_k,U_l)$.
\end{itemize}
The decomposition of a 4-dimensional copula given in \eq{eq_cop_decomp_nonSimp} is not unique.
For a given dimension $d$, admissible decompositions may be represented as vines: sequences of $d-1$ trees, where the nodes of the $(i+1)^{\rm th}$ tree are the edges of the $i^{\rm th}$ tree.
The first tree consists of $d$ nodes, corresponding to the $d$ random variables in the distribution.
Admissible decompositions correspond to \textit{regular} vines,
which are vines satisfying the proximity condition: 
two edges in the $i^{\rm th}$ tree may be joined in the $(i+1)^{\rm th}$ tree only if they share a common node in the $i^{\rm th}$ tree.
For more details, the reader is referred to the references cited above.

In \eq{eq_cop_decomp_nonSimp}, although $c_{1,4;2}$ and $c_{1,3;2}$ 
are bivariate copulas for a fixed value of $u_2$, they are effectively trivariate functions due to explicit dependence on $u_2$.
Similarly, $c_{3,4;1,2}$ represents a bivariate copula that is effectively a quadrivariate function due to its explicit dependence on $u_1$ and $u_2$.
To simplify statistical inference, the \textit{simplifying assumption} ignores explicit dependence on the conditioning variables. Applying the simplifying assumption to \eq{eq_cop_decomp_nonSimp} yields: 
\begin{align}
c_{1234} & (u_1,u_2,u_3,u_4)  =   \nonumber \\
& c_{1,2}(u_1,u_2) \times c_{2,3}(u_2,u_3) \times c_{2,4}(u_2,u_4) \nonumber \\
& \times c_{1,4;2}(C_{1|2}(u_1;u_2),C_{4|2}(u_4;u_2)) \times c_{1,3;2}(C_{1|2}(u_1;u_2),C_{3|2}(u_3;u_2)) \nonumber \\
& \times c_{3,4;1,2}(C_{3|1,2}(u_3;u_1u_2),C_{4|1,2}(u_4; u_1, u_2)) \label{eq_cop_decomp_simp}  \, ,                              
\end{align}
which is equivalent to assuming that 
$(U_{1|2}, U_{4|2}) \perp U_2$,
$(U_{1|2}, U_{3|2}) \perp U_2$,
and $(U_{3|1,2} , U_{4|1,2} ) \perp (U_1,U_2)$, where the symbol ``$\perp$'' denotes independence.

When estimating the copula underlying a high-dimensional dataset, the simplifying assumption makes the problem tractable by mitigating the curse of dimensionality. Under this assumption, estimating a $d$-dimensional copula reduces to the sequential estimation of  
$d(d-1)/2$ bivariate copula densities, such as $c_{1,2}$, $c_{2,3}$, $c_{2,4}$, $c_{1,4;2}$, $c_{1,3;2}$, and $c_{3,4;1,2}$ 
appearing in \eq{eq_cop_decomp_simp}.
These bivariate copulas can be estimated using either parametric or non-parametric methods.
The constraint imposed by the simplifying assumption on the overall multivariate dependence structure is rather intricate, and it permits the representation of a broad class of copulas.
Even when the true copula substantially violates the simplifying assumption, simplified vine-copula estimators have been shown to remain competitive compared with classical multivariate kernel density estimators \cite{nagler_2016}.

For numerical implementation, the R package \texttt{rvinecopulib} \cite{rvinecopulib_package} was used.

\bibliography{mybibfile}

\end{document}